\documentclass[11pt]{article}

\include{setup}

\begin{document}

\thispagestyle{empty}
\def\thefootnote{\fnsymbol{footnote}}
\setcounter{footnote}{1}
\null
\draftdate\hfill PITHA 07/08 \\
\strut\hfill MPP-2007-132 \\
\strut\hfill UWThPh-2007-24\\
\strut\hfill PSI-PR-07-08 \\
\vskip 0cm
\vfill
\begin{center}
  {\Large \boldmath{\bf Radiative corrections to W-boson
      hadroproduction:
      \\[.5em]
      higher-order electroweak and supersymmetric effects}
\par} \vskip 2.5em
{\large
{\sc Silja Brensing$^1$, Stefan Dittmaier$^{2,3}$,
     Michael Kr\"amer$^1$, Alexander M\"uck$^{1,4}$
}\\[2ex]
{\normalsize \it 
$^1$Institut f\"ur Theoretische Physik E, RWTH Aachen, \\ 
D-52056 Aachen, Germany}\\[1ex]
{\normalsize \it 
$^2$Max-Planck-Institut f\"ur Physik (Werner-Heisenberg-Institut), \\
D-80805 M\"unchen, Germany
}\\[1ex]
{\normalsize \it 
$^3$Faculty of Physics, University of Vienna, \\
A-1090 Vienna, Austria     
}\\[1ex]
{\normalsize \it 
$^4$Paul Scherrer Institut, W\"urenlingen und Villigen,\\ 
Ch-5232 Villigen PSI, Switzerland}\\[2ex]
}
\par \vskip 1em
\end{center}\par
\vskip .0cm \vfill {\bf Abstract:} \par 
The high accuracy envisaged for future measurements of W-boson
production at hadron colliders has to be matched by precise
theoretical predictions. We study the impact of electroweak radiative
corrections on W-boson production cross sections and differential
distributions at the Tevatron and at the LHC. In particular, we
include photon-induced processes, which contribute at ${\cal
  O}(\alpha)$, and leading radiative corrections beyond ${\cal
  O}(\alpha)$ in the high-energy Sudakov regime and from multi-photon
final-state radiation.  We furthermore present the calculation of the
complete supersymmetric next-to-leading-order electroweak and QCD
corrections to W-boson hadroproduction within the MSSM. The
supersymmetric corrections turn out to be negligible in the vicinity
of the W resonance in general, reaching the percent level only at
high lepton transverse momentum and for specific choices of the
supersymmetric parameters.

\par
\vskip 1cm
\noindent
\par
\null
\setcounter{page}{0}
\clearpage
\def\thefootnote{\arabic{footnote}}
\setcounter{footnote}{0}

\section{Introduction}
 
The Drell--Yan-like production of W~bosons,
\begin{equation}
\Pp\Pp/\Pp\bar\Pp \to \PW \to \Pl\nu_\Pl \X,
\end{equation}
is one of the cleanest hadron collider processes with a large cross
section at the Tevatron and at the LHC. Measurements near the
\PW~resonance allow for a precise determination of the \PW-boson mass
and yield valuable information on the parton structure of the proton.
Above resonance, the off-shell and high-energy tails of appropriate
distributions are sensitive to the \PW~width and offer the possibility
to search for additional charged gauge bosons $\PW'$.  (See e.g.\ 
\citeres{Gerber:2007xk, Buescher:2006jm} and references therein.)

The high experimental precision envisaged at the Tevatron Run~II and
specifically at the LHC has to be matched by precise theoretical
predictions including QCD and electroweak radiative corrections. The
QCD corrections are completely known up to next-to-next-to-leading
order (NNLO)~\cite{vanNeerven:1991gh} and up to N$^3$LO in the
soft-plus-virtual approximation~\cite{Moch:2005ky}, with a remaining
theoretical error for inclusive cross sections at the percent level or
lower.  The next-to-leading-order (NLO) QCD corrections have been
matched with parton showers \cite{Frixione:2006gn} and combined with a
summation of soft gluon radiation (see e.g.\ \citere{Balazs:1997xd}),
which is particularly important to reliably predict the \PW\ 
transverse-momentum distribution. A theoretical study of the QCD
uncertainties in the determination of the \PW~cross section at hadron
colliders has been presented in \citere{Frixione:2004us}.

In this article we mainly focus on the electroweak corrections. Given,
for example, the anticipated experimental accuracy in the \PW-boson
mass measurement of $10{-}20\MeV$ at the LHC, the inclusion of
electroweak corrections beyond final-state radiation is mandatory as
their omission would induce a systematic error in the \MW\ 
determination of ${\cal O}(10~\mbox{MeV})$ (see e.g.\ 
\citere{Gerber:2007xk}). The complete NLO ${\cal O}(\alpha)$
corrections to the parton process $\Pq\bar{\Pq}' \to \PW \to \Pl
\nu_{\Pl}$ have been calculated by several
groups~\cite{Zykunov:2001mn, Dittmaier:2001ay, Baur:2004ig,
  CarloniCalame:2006zq}.  A tuned comparison of cross sections and
differential distributions has shown good agreement between the
various calculations~\cite{Buttar:2006zd, Gerber:2007xk}.
Photon-induced processes $\gamma \Pq\to \Pl \nu_\Pl \Pq'$, which also
contribute at ${\cal O}(\alpha)$, have been computed recently in
\citeres{DK_LH,Arbuzov:2007kp}.  The particular importance of
final-state photon radiation for the \PW-boson mass determination
demands a treatment that goes beyond ${\cal O}(\alpha)$. Such
multi-photon effects have been studied
in~\citeres{CarloniCalame:2003ux, Placzek:2003zg} and matched to the
NLO ${\cal O}(\alpha)$ calculation in~\citere{CarloniCalame:2006zq}.
First steps towards combining QCD and electroweak higher-order effects
have been taken in \citere{Cao:2004yy}. We note that the ${\cal
  O}(\alpha)$ corrections to hadronic production of on-shell
\PW~bosons at large transverse momenta, $\Pp\Pp/\Pp\bar\Pp\to\PW
+\mbox{jet}$, have recently been presented in
\citere{Kuhn:2007qc} for the Standard Model (SM) and in 
\citere{Gounaris:2007gx} in the
minimal supersymmetric Standard Model (MSSM). For related work on 
electroweak corrections to \PZ-boson hadroproduction we refer to 
\citeres{Buttar:2006zd, Gerber:2007xk} and further references therein.

In this paper we improve on our previous ${\cal O}(\alpha)$
calculation~\cite{Dittmaier:2001ay} for $\Pp\Pp/\Pp\bar\Pp\to\PW\to
\Pl \nu_\Pl \X$ by including photon-induced processes and
multi-photon final-state radiation in the structure-function approach.
We also discuss the impact of leading electroweak effects beyond
${\cal O}(\alpha)$, specifically the Sudakov logarithms that arise in
the high-energy regime, and present the calculation of the complete
supersymmetric ${\cal O}(\alpha)$ electroweak and ${\cal O}(\alpha_{\rm s})$ 
strong corrections within the MSSM. The purpose of the MSSM
calculation is to establish that the impact of virtual supersymmetric
particles on the cross section prediction is small and does not spoil
the status of single-\PW-boson production as one of the
cleanest SM candles at hadron colliders.

The paper is organized as follows. In \refse{se:order_alpha} 
we briefly summarize the calculation of the complete ${\cal O}(\alpha)$ 
corrections as originally presented in \citere{Dittmaier:2001ay}. 
In the following sections, we describe the
inclusion of leading electroweak effects beyond ${\cal O}(\alpha)$, 
the calculation of the photon-induced processes, and the summation of
multiple emission of collinear photons off the final-state lepton.
Numerical results for \PW-boson production at the Tevatron and the LHC
including the complete ${\cal O}(\alpha)$ and ${\cal O}(\alpha_{\rm s})$ 
NLO corrections and the higher-order Sudakov and photon
radiation effects are presented in \refse{se:numres}. 
There, we also compare our results on multi-photon final-state radiation
in the structure-function approach with results
based on the parton-shower approach~\cite{Gerber:2007xk}.
In~\refse{se:mssm} we describe the calculation of the MSSM 
corrections and discuss their numerical impact on the
\PW~cross section and distributions. We conclude in
\refse{se:conclusion}.  Finally, the Appendix provides details
on the scenarios of the supersymmetric models under consideration.

\section{Higher-order electroweak effects}
\label{se:hoew} 

In the next section, we briefly review the calculation of the
complete ${\cal O}(\alpha)$ corrections to the parton process 
$\Pq\bar{\Pq}' \to \PW \to \Pl \nu_{\Pl}$ presented in
\citere{Dittmaier:2001ay} which is now augmented by an extension of 
the dipole subtraction method~\cite{Dittmaier:2008md} that allows
us to calculate non-collinear-safe observables also within the 
subtraction approach. In the subsequent sections, we then
discuss the leading electroweak effects and the choice of couplings
before we present those parts of the calculation that are new and that
have not been discussed in \citere{Dittmaier:2001ay}. We present our
results for the specific case of $\PW^+$~production,
$\Pp\Pp/\Pp\bar\Pp\to\PW^+\to \Pl^+ \nu_\Pl \X$.

\subsection{Electroweak \boldmath{${\cal O}(\alpha)$ corrections}}
\label{se:order_alpha}

We consider the parton process
\begin{equation}
u(p_u) + \bar d(p_d) \;\to\;
\nu_l(k_n) + l^+(k_l) \;\; [+\gamma(k)],
\end{equation}
where $u$ and $d$ generically denote the light up- and down-type quarks,
$u=\Pu,\Pc$ and $d=\Pd,\Ps$. The lepton $l$ represents $l=\Pe,\mu$.
The momenta of the particles are given in brackets, and the 
Mandelstam variables are given by
\begin{equation}
\hat s = (p_u+p_d)^2, \quad
\hat t = (p_d-k_l)^2, \quad
\hat u = (p_u-k_l)^2. 
\end{equation}
We neglect the fermion masses $m_u$, $m_d$, $\Ml$ whenever possible,
i.e.\ we keep these masses only as regulators in the logarithmic mass 
singularities originating from collinear photon emission or exchange.
In lowest order the scattering amplitude reads 
\begin{equation}
\M_0 = \frac{e^2 V^*_{ud}}{2\sw^2} \,
\left[ \bar v_d\gamma^\mu\omega_-u_u\right] \,
\disp\frac{1}{\hat s-\MW^2+\ri\MW\GW(\hat s)} \,
\left[ \bar u_{\nu_l}\gamma_\mu\omega_-v_l\right],
\label{eq:m0}
\end{equation}
with an obvious notation for the Dirac spinors $\bar v_d$, etc., and
the left-handed chirality projector $\omega_-=\frac{1}{2}(1-\gamma_5)$.
The electric unit charge is denoted by $e$, the weak mixing angle is 
fixed by the ratio $\cw^2=1-\sw^2=\MW^2/\MZ^2$ of the W- and Z-boson
masses $\MW$ and $\MZ$, and $V_{ud}$ is the CKM matrix element for
the $ud$ transition. The different choices for the fine-structure constant
$\alpha= e^2/(4 \pi)$ in the squared matrix element are discussed below.

Strictly speaking, Eq.~\refeq{eq:m0} already goes beyond lowest order,
since the W-boson width $\GW(\hat s)$ results from the Dyson summation of
W~self-energy insertions. Renormalizing the W mass and width in the on-shell 
scheme, the Dyson summation directly leads to a {\it running width}, i.e.\
\begin{equation}
\GW(\hat s)\big|_{\mathrm{run}} = \GW \frac{\hat s}{\MW^2}.
\end{equation}
On the other hand, defining W~mass and width from the location of the
pole of the W~propagator (with momentum transfer $p$)
in the complex $p^2$~plane, naturally leads to a {\it constant width}, i.e.\
\begin{equation}
\GW(\hat s)\big|_{\mathrm{const}} = \GW.
\label{eq:fixedwidth}
\end{equation}
In this work, we employ the fixed-width approach \refeq{eq:fixedwidth}. 
This has to be
kept in mind when using the results in precision determinations of 
the W-boson mass from the W~resonance.

The virtual one-loop corrections comprise contributions of the
transverse part of the $W$~self-energy $\Sigma^W_{\mathrm{T}}$, 
corrections to the two $Wdu$ and $W\nu_l l$ vertices, 
box diagrams, and counterterms. 
Our calculation of these corrections is described in 
\citere{Dittmaier:2001ay} in detail. In particular,
the complete expressions for the vertex and box 
corrections are provided in the Appendix of that reference.
For this paper we recalculated the one-loop effects employing the
packages {\tt FeynArts} \cite{Hahn:2000kx}, {\tt FormCalc}, and
{\tt LoopTools}~\cite{Hahn:1998yk}, supplemented by the loop integrals with 
singularities on the W~resonance, which are discussed below.
In both calculations, ultraviolet divergences are treated
in dimensional regularization, and the infrared (IR) singularity is regularized
by an infinitesimal photon mass $m_\gamma$. The actual calculation is
performed in 't~Hooft--Feynman gauge using on-shell renormalization. 

The complete one-loop amplitude $\M_1$ can be expressed in terms of a
correction factor $\delta^{\virt}$ times the lowest-order matrix
element,
\begin{equation}
\M_1 = \delta^{\virt} \M_0.
\label{eq:m1}
\end{equation}
Thus, in ${\cal O}(\alpha)$ the squared matrix element reads
\begin{equation}
|\M_0+\M_1|^2 = (1+2\Re\{\delta^{\virt}\}) |\M_0|^2 + \dots,
\label{eq:dm2virt}
\end{equation}
so that the Breit--Wigner factors are completely contained in the
lowest-order factor $|\M_0|^2$. Note that the Dyson-summed imaginary
part of the W~self-energy, which appears as $\GW(\hat s)$ in $\M_0$,
is not double-counted, since only the real part of
$\Sigma^W_{\mathrm{T}}$ enters $\Re\{\delta^{\virt}\}$ in ${\cal
O}(\alpha)$.
Despite the separation of the resonance pole $(\hat s-\MW^2)^{-1}$ 
from the correction factor $\delta^{\virt}$ in \refeq{eq:m1},
$\delta^{\virt}$ still contains logarithms $\ln(\hat s-\MW^2+\ri\epsilon)$ 
that are singular on resonance. Since these singularities would be cured 
by a Dyson summation of the $W$~self-energy inside the loop diagrams, we
substitute
\begin{equation}
\ln(\hat s-\MW^2+\ri\epsilon) \;\to\; \ln(\hat s-\MW^2+\ri\MW\GW) 
\label{eq:onshelllogs}
\end{equation}
with a fixed width everywhere. The substitution \refeq{eq:onshelllogs}
does not disturb the gauge-invariance properties of the one-loop 
amplitude $\M_1$.

In \citere{Dittmaier:2001ay} (Section~2.2) three different
input-parameter schemes have been specified for the choice of the
electromagnetic coupling constant $\alpha$ that is taken as SM input
together with the particle masses. According to the choice of
$\alpha$, the schemes are called ``$\alpha(0)$'', ``$\alpha(\MZ^2)$'',
and ``$\GF$'' schemes, where $\alpha$ is set to
$\alpha_{\GF}=\sqrt{2}\GF\MW^2(1-\MW^2/\MZ^2)/\pi$ in the $\GF$ scheme.  
In the $\alpha(0)$ scheme, the charge 
renormalization constant $\delta Z_e$ in the counterterm 
$\delta_{Wff'}^{\mathrm{ct}}$ of the $Wff'$ vertex 
contains logarithms of the light-fermion masses
which are related to the running of the electromagnetic coupling
$\alpha(Q^2)$ from $Q=0$ to a high-energy scale. In the 
$\alpha(\MZ^2)$ scheme the counterterm changes to
\begin{equation}
\delta_{Wff'}^{\mathrm{ct}}\big|_{\alpha(\MZ^2)} =
\delta_{Wff'}^{\mathrm{ct}}\big|_{\alpha(0)} -
\frac{1}{2}\Delta\alpha(\MZ^2),
\end{equation}
where 
\begin{equation}
\Delta\alpha(Q^2) = \Pi^{AA}_{f\ne \Pt}(0)-\Re\{\Pi^{AA}_{f\ne \Pt}(Q^2)\},
\end{equation}
\begin{sloppypar}
\noindent
with $\Pi^{AA}_{f\ne \Pt}$ denoting the photonic vacuum polarization
induced by all fermions other than the top quark. 
In contrast to the $\alpha(0)$ scheme the 
coun\-ter\-term $\delta_{Wff'}^{\mathrm{ct}}\big|_{\alpha(\MZ^2)}$
does not involve light quark masses, i.e.\ the large logarithmic
corrections are absorbed in the lowest order
by using the appropriate numerical value for $\alpha(\MZ^2)$.
In the $\GF$ scheme, the transition from $\alpha(0)$ to $\GF$ is ruled 
by the quantity $\Delta r$ which is deduced from muon decay,
\end{sloppypar}
\begin{equation}
\alpha_{\GF}=\frac{\sqrt{2}\GF\MW^2\sw^2}{\pi}
=\alpha(0)(1+\Delta r) \;+\; {\cal O}(\alpha^3).
\end{equation}
Therefore, the counterterm $\delta_{Wff'}^{\mathrm{ct}}$ reads
\begin{equation}
\delta_{Wff'}^{\mathrm{ct}}\big|_{\GF} =
\delta_{Wff'}^{\mathrm{ct}}\big|_{\alpha(0)} - \frac{1}{2}\Delta r.
\end{equation}
Since $\Delta\alpha(\MZ^2)$ is implicitly contained in $\Delta r$, the
large fermion-mass logarithms are also absorbed in the lowest
order in the $\GF$ scheme. 
The different input-parameter schemes are further discussed in 
Section~\ref{se:ips}.

The calculation of the real-photonic corrections is described in
\citere{Dittmaier:2001ay} in detail, both for a running and a fixed
W-boson width. In order to respect electromagnetic gauge invariance,
the coupling of 
the photon to \PW~bosons has to be adapted when a running width is used.
In the fixed-width approach, as used in this paper, no modification of
couplings is needed.

The helicity amplitudes for the radiative process are explicitly 
given in Section~2.4 of \citere{Dittmaier:2001ay}.
The phase-space integral over the real-photonic matrix elements 
diverges in the soft and collinear regions logarithmically
if the photon and fermion masses are set to zero. 
To properly combine the soft and collinear singularities 
with the corresponding virtual corrections
three different methods are applied:
two variants of phase-space slicing and the 
dipole subtraction method.

The dipole subtraction approach as formulated 
in \citere{Dittmaier:1999mb} can only be applied for 
collinear-safe observables, i.e. observables for which 
selection cuts are blind to the distribution of momenta 
in collinear lepton--photon configurations. This can be achieved 
by ``photon recombination'', where leptons and sufficiently 
collinear photons are treated as one quasi-particle
(see also \refse{se:cuts}). For these
observables the KLN theorem \cite{Kinoshita:1962ur} guarantees 
that logarithms of the fermion masses are absent in the
corrections. For muons in the final state it is, however, 
experimentally possible to separate collinear photons from the
lepton, i.e.\ to observe so-called ``bare'' muons. Hence, the
resulting cross sections are not collinear safe and the
corresponding collinear singularities show up
as logarithms of the small lepton masses.

In \citere{Dittmaier:2001ay}, only the slicing variants were able to deal
with bare leptons, while the application of the subtraction approach 
was still restricted to collinear-safe observables.
In this work, we employ an extension~\cite{Dittmaier:2008md} 
of the subtraction formalism which allows one
to calculate cross sections for bare leptons, i.e.\ cross sections
defined without any photon recombination. 
The respective results of the slicing and subtraction 
methods are in good numerical agreement
both for the inclusive and the bare muon case. 

\subsection{Leading electroweak effects and choice of couplings}
\label{se:ips}

As described in \citere{Dittmaier:2001ay}, the relative corrections in
the various input-parameter schemes 
differ by constant contributions proportional to
$\De\alpha(\MZ^2)\approx6\%$ and $\De r\approx3\%$, 
which quantify the running of the electromagnetic coupling from
$Q^2=0$ to $Q^2=\MZ^2$ for $\alpha(\MZ^2)$ and the radiative corrections 
to muon decay for the $\GF$ scheme,
respectively. The bulk of $\De r$ is contained in
$\De\alpha(\MZ^2)-\cw^2\De\rho/\sw^2$, where $\De\rho\approx1\%$ is the
universal correction $\propto\GF\Mt^2$ to the $\rho$ parameter.

The $\GF$ scheme is distinguished from the two other schemes because
the corrections to charged-current four-fermion processes do
not contain large contributions from $\De\alpha(\MZ^2)$ or $\De\rho$
anymore, i.e.\ these universal renormalization effects are completely
absorbed into  the leading-order (LO) 
amplitude. For the ${\cal O}(\alpha)$ 
corrections this property has already been pointed out in
\citere{Dittmaier:2001ay}, but it also holds at ${\cal O}(\alpha^2)$.
More precisely, in the $\GF$ scheme no contributions proportional to
$\De\alpha(\MZ^2)^n$ (for any positive integer $n$), proportional to
$\De\alpha(\MZ^2)\De\rho$, and proportional to $\De\rho^2$ appear.
Based on the arguments given in \citere{Consoli:1989fg} this feature
was explicitly worked out in \citere{Diener:2005me} (Section~3) for
the related process of charged-current neutrino deep-inelastic
scattering; these results obviously apply also to charged-current
Drell--Yan scattering via crossing symmetry.

\looseness-1
Before concluding that the theoretical uncertainties in the $\GF$
scheme from missing corrections beyond ${\cal O}(\alpha)$ are smaller
than in the two other schemes, we shall inspect other known universal
dominant corrections at ${\cal O}(\alpha)$.  Besides the
renormalization effects discussed above, the dominant ${\cal
  O}(\alpha)$ corrections, up to moderate parton scattering energies,
are due to final-state radiation off the charged lepton, at least for 
bare leptons where 
the enhancement by the large mass logarithm $\propto\alpha\ln \Ml$ is
present.  The inclusion of these contributions beyond ${\cal
  O}(\alpha)$, which are due to collinear multi-photon emission, is
described in \refse{se:Multi-photon} below. Here we merely point out
that the appropriate coupling constant $\alpha$ entering the relative
correction is $\alpha(0)$, because it accounts for the emission of
photons with $Q^2=0$. Thus, when adopting an input-parameter scheme
other than the $\alpha(0)$ scheme, one should nevertheless 
use $\alpha(0)$ to multiply the dominating universal lepton-mass
logarithms. Our specific implementation of this procedure is described
in \refse{se:Multi-photon}.

At high parton scattering energies and high transverse lepton momenta,
electroweak corrections are dominated by soft and/or collinear
gauge-boson exchange. The soft effects, known as Sudakov logarithms,
induce powers of $\alpha\ln^2(\hat s/\MW^2)$, with subleading
soft/collinear contributions involving lower powers in the logarithm.
In ${\cal O}(\alpha)$, these enhanced electroweak effects drive the
relative corrections to $\approx -30\%$ at lepton transverse momenta of
about $1\TeV$ at the LHC \cite{Dittmaier:2001ay}. It is therefore
desirable to control this kind of corrections beyond ${\cal
  O}(\alpha)$.  We elaborate more on this issue in the next section.
Here we just point out that among the considered input-parameter 
schemes the $\GF$ scheme should be 
appropriate to fix $\alpha$ for the leading high-energy logarithms,
which are of weak origin. 

Following the above arguments, we employ the $\GF$ scheme in this
work, modified only by the change of $\alpha_{\GF}$ to $\alpha(0)$ in
the leading part of final-state radiation. This procedure is expected 
to be most robust with respect to further corrections beyond 
${\cal O}(\alpha)$.

\subsection{Leading weak corrections in the Sudakov regime}
\label{se:sudakov}

For single-\PW\ production at large lepton transverse momenta or \PW\ 
transverse masses, the parton kinematics is restricted to the Sudakov
regime, characterized by large Mandelstam parameters $\hat{s}$,
$|\hat{t}|$, $|\hat{u}| \gg \MW^2$. The structure of electroweak
corrections beyond ${\cal O}(\alpha)$ in this high-energy regime has
been investigated in some detail by several groups in recent years
(see e.g.\ \citeres{Fadin:1999bq, Ciafaloni:2000df, Hori:2000tm,
  Melles:2001dh, Beenakker:2001kf, Denner:2003wi, Jantzen:2005xi,
  Denner:2006jr} and references therein).

As described for example in \citeres{Denner:2003wi,Denner:2006jr}, the
leading electroweak logarithmic corrections, which are enhanced by
large factors $L=\ln(\hat s/\MW^2)$, can be divided into an
SU(2)$\times$U(1)-symmetric part, an electromagnetic part, and a
subleading part induced by the mass difference between $\PW$ and
\PZ~bosons.  The last part does not contribute to corrections
$\propto(\alpha L^2)^n$ and is neglected in the following.  The
leading (Sudakov) logarithms $\propto(\alpha L^2)^n$ 
of electromagnetic origin cancel between
virtual and real (soft) bremsstrahlung corrections; for the subleading
logarithms such cancellations should strongly depend on the observable
under consideration.  The only source of leading logarithms is, 
thus, the symmetric electroweak (sew)
part, which can be characterized by comprising \PW~bosons, \PZ~bosons,
and photons of a common mass $\MW$.  Using this mass assignment, the
one-loop correction $\de^{(1)}_{\mathrm{sew}}$ to the squared
amplitude can be obtained by expanding the full result for the
virtual correction $\de^{\virt}$ (given in Appendix~A of
\citere{Dittmaier:2001ay}) for large $\hat s$, $|\hat{t}|$, 
$|\hat{u}| \gg \MW^2$. The explicit result can be written as
\begin{equation}
\de^{(1)}_{\mathrm{sew}} =
\frac{\alpha}{2\pi}\left\{
-L^2 C^{\mathrm{sew}}_{1,\mathrm{CC}}
+L C^{\mathrm{ad}}_{1,\mathrm{CC}} \right\}
\label{eq:Sud1loop}
\end{equation}
with factors
\begin{equation}
C^{\mathrm{sew}}_{1,\mathrm{CC}} =
\frac{3}{2\sw^2}
+\frac{Y_{\Pu_{\mathrm{L}}}^2+Y_{\nu_{\Pl,\mathrm{L}}}^2}{4\cw^2},
\qquad
C^{\mathrm{ad}}_{1,\mathrm{CC}} =
-\frac{2}{\sw^2} \biggl[\ln\biggl(\frac{-\hat{t}}{\hat s}\biggr)
+\ln\biggl(\frac{-\hat{u}}{\hat s}\biggr)\biggr]
+\frac{2}{\cw^2}Y_{\Pu_{\mathrm{L}}}Y_{\nu_{\Pl,\mathrm{L}}}
\ln\biggl(\frac{\hat{u}}{\hat{t}}\biggr),
\end{equation}
which have been introduced in Section~8.4.2 of \citere{Denner:2006jr}.
Here $Y_{\Pu_{\mathrm{L}}}=1/3$ and $Y_{\nu_{\Pl,\mathrm{L}}}=-1$ are
the weak hypercharges of the corresponding left-handed particles. In
Eq.~\refeq{eq:Sud1loop} we did not only include the leading Sudakov
logarithms $\propto \alpha L^2$, but also the related
``angular-dependent'' contributions $\propto \alpha L \ln(-\hat{t}/\hat s)$
or $ \alpha L \ln(-\hat{u}/\hat s)$.  Our explicit ${\cal O}(\alpha)$ result
is in agreement with the general results presented in
\citeres{Denner:2003wi,Denner:2006jr}, where the corresponding
corrections are also given at the two-loop level.  These ${\cal
  O}(\alpha^2)$ corrections can be obtained from the ${\cal
  O}(\alpha)$ result by an appropriate exponentiation
\cite{Melles:2001dh}.  For the leading ``sew'' corrections (including
$\alpha^2L^4$, $\alpha^2L^3\ln(-\hat{t}/\hat{s})$, and 
$\alpha^2L^3 \ln(-\hat{u}/\hat s)$
terms) this exponentiation simply reads \cite{Denner:2006jr}
\begin{equation}
|\M|^2 \sim |\M_0|^2\exp\left\{\de^{(1)}_{\mathrm{sew}}\right\}
= |\M_0|^2\left( 1+\de^{(1)}_{\mathrm{sew}}+\de^{(2)}_{\mathrm{sew}}
+\dots\right)
\end{equation}
with
\begin{equation}
\de^{(2)}_{\mathrm{sew}} =
\biggl(\frac{\alpha}{2\pi}\biggr)^2\biggl\{
\frac{1}{2}L^4 (C^{\mathrm{sew}}_{1,\mathrm{CC}})^2
-L^3 C^{\mathrm{sew}}_{1,\mathrm{CC}}
C^{\mathrm{ad}}_{1,\mathrm{CC}} \biggr\}.
\label{eq:Sud2loop}
\end{equation}

However, in the case of neutral-current-induced fermion--antifermion
scattering processes it was observed \cite{Jantzen:2005xi} that large
cancellations take place between leading and subleading logarithms.
In view of this uncertainty, we do not include the two-loop
high-energy logarithms in our full predictions. Instead, we evaluate
the leading two-loop part $\de^{(2)}_{\mathrm{sew}}$ as a measure of
missing electroweak corrections beyond ${\cal O}(\alpha)$ in the
high-energy Sudakov regime.

Moreover, since the electroweak high-energy logarithmic corrections
are associated with virtual soft and/or collinear weak-boson or photon
exchange, they all have counterparts in real weak-boson or photon
emission processes which can partially cancel (but not completely, see
\citere{Ciafaloni:2000df}) the large negative corrections. To which
extent the cancellation occurs depends on the experimental
possibilities to separate final states with or without weak bosons or
photons. This issue is discussed for example in
\citeres{Ciafaloni:2006qu,Baur:2006sn}. The numerical analysis
presented in \citere{Baur:2006sn} demonstrates the effect of real
weak-boson emission in the distributions in the transverse lepton
momentum $p_{\mathrm{T},\Pl}$ and in the transverse mass 
$M_{\mathrm{T},\nu_{\Pl} \Pl}$ of the W~boson (as e.g.\ defined in
\refse{se:numreslhc} below). 
For $\PW^+$ production at the LHC, at $M_{\mathrm{T},\nu_{\Pl} \Pl}=2\TeV$
the electroweak corrections are reduced from about $-26\%$ to $-23\%$
by weak-boson emission. At $p_{\mathrm{T},\Pl}=1\TeV$
the corresponding reduction from about $-28\%$ to $-17\%$ is much larger,
however, the bulk of these emission effects is not due to soft/collinear 
weak-boson emission, but due to recoil effects in real $\PW^+\PW^-$ 
and $\PW^+\PZ$ production. This explicitly illustrates the sensitivity
of weak-boson emission effects to the details of experimental
event selection, in particular, how single-W production is separated
from di-boson production.

\subsection{Photon-induced processes}
\label{se:Photon-induced}

The ${\cal O}(\alpha)$ corrections to the parton cross section
$\Pq\bar{\Pq}' \to \PW^+ \to \Pl^+ \nu_{\Pl}$ contain
collinear singularities from photon radiation off the initial-state
quarks which are absorbed by mass factorization
\cite{Dittmaier:2001ay, Baur:2004ig, CarloniCalame:2006zq}. For a
complete and theoretically consistent analysis, the absorption of the
${\cal O}(\alpha)$ collinear singularities into quark distributions
has to be complemented by both the inclusion of ${\cal O}(\alpha)$
corrections to the parton distribution functions (PDFs) and by the
inclusion of the ${\cal O}(\alpha)$ partonic subprocesses
$\gamma \, \Pu \to \Pl^+ \nu_\Pl \Pd$ and $\gamma \, \Pdbar \to \Pl^+
\nu_\Pl \Pubar$, which were first calculated in \citere{DK_LH}.
At the time of our previous study
\cite{Dittmaier:2001ay} a complete analysis of ${\cal O}(\alpha)$
corrections to PDFs, which involves ${\cal O}(\alpha)$ corrections to
the DGLAP evolution and to the fit of experimental data, was not
available. Meanwhile, 
the MRSTQED2004~\cite{Martin:2004dh} PDF parametrization,
however, includes these ${\cal O}(\alpha)$ corrections and provides a
photon density within protons and antiprotons. It is thus now possible
to perform a complete ${\cal O}(\alpha)$ analysis and to properly
include the photon-induced subprocesses in the hadronic cross-section
prediction.

The collinear photon splitting into two massless quarks in the
subprocesses $\gamma \, \Pu \to \Pl^+ \nu_\Pl \Pd$ and $\gamma \,
\Pdbar \to \Pl^+ \nu_\Pl \Pubar$ also leads to a mass singularity. As
usual, this divergence is removed by mass factorization. Including
both the divergences from collinear photon splitting and the
divergence due to collinear photon emission from initial-state quarks
in $\Pq\bar{\Pq}' \to \Pl^+ \nu_{\Pl} \gamma$, mass
factorization implies a redefinition of the quark densities according
to \cite{Diener:2005me} 
\begin{equation}
\begin{split}
  f_{\Pq}(x) \to f_{\Pq}(x,\mu_F) & - \int_{x}^{1} \, \frac{dz}{z} \,
  \,
  f_\Pq \left(\frac{x}{z},\mu_F \right) \, Q_{\Pq}^2 \, \frac{\alpha}{2\pi} \\
  & \quad \quad \quad \! \times \!  \left\{ \ln \left(
      \frac{\mu_F^2}{\Mq^2} \right) \Bigl[ P_{\Pq\Pq}(z) \Bigr]_+ -
    \Bigl[ P_{\Pq\Pq}(z) \left( 2 \ln(1-z) +1  \right) \Bigr]_+ + C_{\Pq\Pq}(z) \right\} \, \\
  & - \int_{x}^{1} \, \frac{dz}{z} \, f_\gamma \left(\frac{x}{z},\mu_F
  \right) \, 3 \, Q_{\Pq}^2 \, \frac{\alpha}{2 \pi} \, \left\{ \ln
    \left( \frac{\mu_F^2}{\Mq^2} \right) P_{\Pq\gamma}(z) \, +
    C_{\Pq\gamma}(z) \right\} \, ,
\end{split}
\end{equation}
where $Q_{\Pq}$ is the electric quark charge, $\Mq$ is the small quark
mass used as a regulator, and $\mu_F$ denotes the QED factorization
scale which is identified with the QCD factorization scale.  The
factor 3 in the second line stems from the splitting of the photon
into $\Pq \Pqbar$~pairs of different color. Furthermore,
\begin{equation}
P_{\Pq\Pq}(z) = \frac{1+z^2}{1-z}, \qquad 
P_{\Pq\gamma}(z) =  z^2 + (1-z)^2 
\end{equation}
are the quark and photon splitting functions, respectively, and
$C_{\Pq\Pq}$, $C_{\Pq\gamma}$ the coefficient functions specifying the
factorization scheme. Following standard QCD terminology one
distinguishes \MSbar\ and DIS schemes defined by
\begin{eqnarray}
C^{\overline{\mathrm{MS}}}_{\Pq\Pq}(z) & = & C^{\overline{\mathrm{MS}}}_{\Pq\gamma}(z)  = 0,
\nonumber \\
C^{\rm DIS}_{\Pq\Pq}(z) & = & \left[ P_{\Pq\Pq}(z) \left( \ln \frac{1-z}{z} -
                   \frac{3}{4} \right) + \frac{9+5 z}{4} \right]_+ \, \, ,\\
C^{\rm DIS}_{\Pq\gamma}(z) & = & P_{\Pq\gamma}(z) \ln \frac{1-z}{z} -
                     8 z^2 + 8 z -1 \, \, . \nonumber
\end{eqnarray}

In our numerical analysis we employ the MRSTQED2004 parton
distribution functions.  Note that photon radiation off incoming
quarks was ignored in the $F_2$ fit to HERA data in the MRSTQED2004
PDF determination. Therefore, the MRSTQED2004 PDFs are defined in the
DIS scheme for the factorization of QED effects, i.e.\ {\it not} in
the $\overline{\mathrm{MS}}$ scheme as frequently done in the past
(see also \citere{Diener:2005me}). For the factorization of QCD
effects the PDFs are defined in the $\overline{\mathrm{MS}}$ scheme 
as usual.

To extract the collinear divergence from the squared matrix element
for the photon-induced processes we use an
extension~\cite{Dittmaier:2008md} of the dipole subtraction
technique, which has been formulated to treat the collinear splitting
of photons into light fermions $\gamma \to \Pf\bar{\Pf}$.

\subsection{Multi-photon final-state radiation}
\label{se:Multi-photon}

The emission of photons collinear to the outgoing charged lepton leads
to corrections that are enhanced by large logarithms of the form
$\alpha\ln(\Ml^2/Q^2)$ with $Q$ denoting a characteristic scale  
of the process. The
KLN theorem \cite{Kinoshita:1962ur} guarantees that these logarithms
cancel if photons collinear to the lepton are treated fully
inclusively.  However, since we apply a phase-space cut on the
momentum of the outgoing lepton, contributions enhanced by these
logarithms survive if the momentum of the bare lepton is considered,
i.e.\ if no photon recombination is performed. While the concept of a
bare lepton is not realistic for electrons, it is phenomenologically
relevant for muon final states.

The first-order logarithm $\alpha\ln(\Ml^2/Q^2)$ is, of course,
contained in the full ${\cal O}(\alpha)$ correction, so that $Q$ is
unambiguously fixed in this order.  However, it is desirable to
control the logarithmically enhanced corrections beyond 
${\cal O}(\alpha)$. This can be done in the so-called structure-function
approach \cite{Kuraev:1985hb}, where these logarithms are derived from
the universal factorization of the related mass singularity.  The
incorporation of the mass-singular logarithms takes the form of a
convolution integral over the LO cross section
$\sigma_{0}$,
\begin{equation}
  \sigma_{\llog\FSR} =
  \int \rd\sigma_{0}(p_\Pu,p_\Pd;k_{\nu_\Pl},k_\Pl)
  \int^1_0 \rd z \, \Gamma_{\Pl\Pl}^{\llog}(z,Q^2) \,
  \Theta_{\cut}(z k_\Pl),
\label{eq:llfsr}
\end{equation}
where the step function $\Theta_{\cut}$ is equal to 1 if the event
passes the cut on the rescaled lepton momentum $z k_\Pl$ and 0
otherwise.  The variable $z$ is the momentum fraction describing the
lepton energy loss by collinear photon emission. Note that in contrast
to the parton-shower approaches to photon radiation (see e.g.\ 
\citeres{CarloniCalame:2003ux, Placzek:2003zg, CarloniCalame:2006zq}),
the structure-function approach neglects the photon momenta transverse
to the lepton momentum. 

For the structure function $\Gamma_{\Pl\Pl}^{\llog}(z,Q^2)$ we take
into account terms up to ${\cal O}(\al^3)$ improved by the well-known
exponentiation of the soft-photonic parts \cite{Kuraev:1985hb},
\newcommand{\betal}{\be_\Pl}
\begin{eqnarray}  
  \Gamma_{\Pl\Pl}^{\llog}(z,Q^2) &=&    
    \frac{\exp\left(-\frac{1}{2}\betal\gamma_{\rE} +
        \frac{3}{8}\betal\right)}
{\Gamma\left(1+\frac{1}{2}\betal\right)}
    \frac{\betal}{2} (1-z)^{\frac{\betal}{2}-1} - \frac{\betal}{4}(1+z) 
\nn\\
&&  {} - \frac{\betal^2}{32} \biggl\{ \frac{1+3z^2}{1-z}\ln(z)
    + 4(1+z)\ln(1-z) + 5 + z \biggr\}
\nn\\
&&  {} - \frac{\betal^3}{384}\biggl\{
      (1+z)\left[6\Li(z)+12\ln^2(1-z)-3\pi^2\right] 
\nn\\
&& \quad\quad {}
+\frac{1}{1-z}\biggl[ \frac{3}{2}(1+8z+3z^2)\ln(z) 
+6(z+5)(1-z)\ln(1-z)
\nn\\
&& \quad\quad\quad {}
+12(1+z^2)\ln(z)\ln(1-z)-\frac{1}{2}(1+7z^2)\ln^2(z)
\nn\\
&& \quad\quad\quad  {}
+\frac{1}{4}(39-24z-15z^2)\biggr] \biggr\} \, ,
\label{eq:GammaFSR}
\end{eqnarray}  
with $\gamma_E$ and $\Gamma(y)$ denoting Euler's constant and
the Gamma function, respectively. The large logarithm is 
contained in the variable
\begin{equation}
\betal = \frac{2\alpha(0)}{\pi} 
\left[\ln\biggl(\frac{Q^2}{\Ml^2}\biggr)-1\right].
\end{equation}
Here, $\alpha(0)$ is the fine-structure constant defined in the
Thomson limit.  The parts solely proportional to a power of $\betal$
correspond to collinear (multi-)photon emission off the lepton, the
exponential factor describes resummed soft-photonic effects. The
non-logarithmic term ``$-1$'' in $\betal$ accounts for a non-singular
universal soft-photonic correction.

Technically, we add the cross section \refeq{eq:llfsr} to the one-loop
result and subtract the LO and one-loop contributions
\begin{equation}
  \sigma_{\llog^1\FSR} =
  \int \rd\sigma_{0}(p_\Pu,p_\Pd;k_{\nu_\Pl},k_\Pl) \int^1_0 \rd z \, 
  \left[\delta(1-z)+\Gamma_{\Pl\Pl}^{\llog,1}(z,Q^2)\right] \,
  \Theta_{\cut}(z k_\Pl),
\end{equation}
contained in \refeq{eq:llfsr} in order to avoid double counting. 
The one-loop contribution to the structure function reads
\begin{eqnarray}  
\label{eq:LLll}
  \Gamma_{\Pl\Pl}^{\llog,1}(z,Q^2) &=&
  \frac{\betal}{4} \left(\frac{1+z^2}{1-z}\right)_+ .
\end{eqnarray}  

More precisely, we adapt the value of $\alpha$ in
$\Gamma_{\Pl\Pl}^{\llog,1}(z,Q^2)$ to the chosen input parameter
scheme. Thereby, we introduce an additional higher-order
contribution $(\alpha(0)-\alpha) \ln(\Ml^2/Q^2)$ so that the $\alpha
\ln(\Ml^2/Q^2)$ contribution to the full ${\cal O}(\alpha)$ correction
is subtracted exactly. Hence, the procedure of adding higher-order
final-state radiation changes also the value of $\alpha$ in the
$\alpha\ln(\Ml^2/Q^2)$ term to $\alpha(0)$ which is the appropriate
coupling for real-photonic effects.

The uncertainty that is connected with the choice of $Q^2$ enters in
${\cal O}(\alpha^2)$, since all $\Oa$ corrections, including constant
terms, are taken into account.  As default we choose the value
\begin{equation}
Q = \xi\sqrt{\hat s}
\label{eq:FSR_scale}
\end{equation}
with $\xi=1$.  In order to quantify the scale uncertainty, we vary
$\xi$ between $1/3$ and $3$.

\subsection{Numerical results}
\label{se:numres}

\subsubsection{Input parameters and setup}
\label{se:SMinput}

The relevant SM input parameters are
\begin{equation}\arraycolsep 2pt
\begin{array}[b]{lcllcllcl}
\GF & = & 1.16637 \times 10^{-5} \GeV^{-2}, \quad&
\alpha(0) &=& 1/137.03599911, &
\alpha_{\mathrm{s}}(\MZ) &=& 0.1189,\\
\MW & = & 80.403\GeV, &
\Gamma_\PW & = & 2.141\GeV, \\
\MZ & = & 91.1876\GeV, &
M_\PH & = & 115\GeV, \\
m_\Pe & = & 0.51099892\MeV, &
m_\mu &=& 105.658369\MeV,\quad &
m_\tau &=& 1.77699\GeV, \\
m_\Pu & = & 66\MeV, &
m_\Pc & = & 1.2\GeV, &
m_\Pt & = & 174.2\;\GeV,
\\
m_\Pd & = & 66\MeV, &
m_\Ps & = & 150\MeV, &
m_\Pb & = & 4.6\GeV,
\\
|V_{\Pu\Pd}| & = & 0.974, &
|V_{\Pu\Ps}| & = & 0.227, \\
|V_{\Pc\Pd}| & = & 0.227, &
|V_{\Pc\Ps}| & = & 0.974,
\end{array}
\label{eq:SMpar}
\end{equation}
which essentially follow \citere{Yao:2006px}.  The masses of the light
quarks are adjusted to reproduce the hadronic contribution to the
photonic vacuum polarization of \citere{Jegerlehner:2001ca}. The CKM
matrix is included via global factors in the partonic cross sections
for the different initial-state quark flavours.
Within loops the CKM matrix is set to unity.

As explained in \refse{se:ips}, we adopt the $\GF$ scheme (up to the
modification of $\alpha$ in the final-state radiation), where the
electromagnetic coupling $\alpha$ is set to $\alpha_{\GF}$. The charge
renormalization constant, which contains logarithms of the fermion
masses, drops out in the $\GF$ scheme so that our results are
practically independent of the light-quark masses. We keep finite
light-quark masses in closed fermion loops, their numerical impact is
however extremely small. The \PW-boson resonance is treated with a
fixed width without any running effects.

The ${\cal O}(\alpha)$-improved MRSTQED2004 set of
PDFs~\cite{Martin:2004dh} is used throughout. 
The QCD and QED factorization scales are
identified and set to the \PW-boson mass $\MW$.

\subsubsection{Phase-space cuts and event selection}
\label{se:cuts}

For the experimental identification of the process
$\Pp\Pp/\Pp\bar\Pp\to\PW^+\to \Pl^+\nu_\Pl \X$ we impose the set of
phase-space cuts
\begin{equation}
p_{\mathrm{T},\Pl}>25\GeV, \qquad
\dsl{p}_\mathrm{T}>25\GeV, \qquad
|\eta_\Pl|<2.5,
\label{eq:lcuts}
\end{equation}
where $p_{\mathrm{T},\Pl}$ and $\eta_\Pl$ are the transverse momentum
and the rapidity of the charged lepton $\Pl^+$, respectively, and
$\dsl{p}_\mathrm{T}=p_{\mathrm{T},\nu_\Pl}$ is the missing
transverse momentum carried away by the neutrino. Note that compared
to our previous study \cite{Dittmaier:2001ay} we have changed the
$\eta_\Pl$-cut to $|\eta_\Pl|<2.5$, which is a more realistic estimate
of the experimental charged-lepton coverage at the LHC.  The
identification cuts are not collinear safe with respect to the lepton
momentum, so that observables in general receive corrections that
involve large lepton-mass logarithms of the form $\alpha\ln(\Ml/\MW)$.
This is due to the fact that photons within a small collinear cone
around the charged-lepton momentum are not treated inclusively, i.e.\ 
the cuts assume a perfect isolation of photons from the charged
lepton. While this is (more or less) achievable for muon final states,
it is not realistic for electrons.  In order to be closer to the
experimental situation for electrons, the following photon
recombination procedure is applied:
\begin{enumerate}
\item Photons with a rapidity $|\eta_\gamma| > 3$, which are close to
  the beams, are considered part of the proton remnant and are not
  recombined with the lepton.\footnote{Note that collinear safety
    requires that the $|\eta_\gamma|$ cut must be larger than the
    lepton identification cut on $|\eta_\Pl|$ to avoid events where an
    almost collinear lepton--photon pair is not recombined because
    $|\eta_\Pl| \lsim 2.5$ and $|\eta_\gamma| \gsim 2.5$. It turns
    out, however, that the numerical difference between choosing
    $|\eta_\gamma| > 3$ and $|\eta_\gamma| > 2.5$ is negligible.}
\item If the photon survived the first step, and if the resolution
  $R_{\Pl\gamma} = \sqrt{(\eta_\Pl-\eta_\gamma)^2 + \phi_{\Pl\gamma}^2}$ is
  smaller than 0.1 (with $\phi_{\Pl\gamma}$ denoting the angle between
  lepton and photon in the transverse plane), then the photon is
  recombined with the charged lepton, i.e.\ the momenta of the photon
  and of the lepton $\Pl$ are added and associated with the momentum of
  $\Pl$, and the photon is discarded.
\item Finally, all events are discarded in which the resulting
  momentum of the charged lepton does not pass the cuts given in
  (\ref{eq:lcuts}).
\end{enumerate}
While the electroweak corrections differ for final-state electrons and
muons without photon recombination, the corrections become universal
in the presence of photon recombination, since the lepton-mass
logarithms cancel in this case, in accordance with the KLN theorem.
Numerical results are presented for photon recombination and for bare
muons.

\subsubsection{\boldmath{Cross sections and distributions for 
  $\Pp\Pp\to\PW^+\to \Pl^+ \nu_\Pl \X$ at the LHC}}
\label{se:numreslhc}

We first consider $\PW^+$~production at the LHC, i.e.\ a
\Pp\Pp~initial state with a centre-of-mass (CM) energy of $\sqrt{s}=14\TeV$.

In Tables~\ref{ta:pt} and \ref{ta:mt} we present the LO cross section
$\sigma_0$ and various types of electroweak corrections $\de$, defined
relative to the LO cross section by $\sigma =
\sigma_0\times\left(1+\de\right)$. The results are shown for different
ranges in the transverse momentum of the charged lepton,
$p_{\mathrm{T},\Pl}$, and in the transverse mass of the two final-state
leptons, $M_{\mathrm{T},\nu_\Pl \Pl} =
\sqrt{2p_{\mathrm{T},\Pl}\dsl{p}_{\mathrm{T}}(1-\cos\phi_{\nu_\Pl
    \Pl})}$, where $\phi_{\nu_\Pl \Pl}$ is the angle between the
lepton and the missing momentum in the transverse plane.

\btab 
\bce  \begin{tabular}{c|cccccc}
 \multicolumn{7}{c}{$\Pp\Pp\to \Pl^+ \nu_\Pl \X$ at $\sqrt{s}=14\TeV$}
 \\  \hline$p_{\mathrm{T},\Pl}/\mathrm{GeV}$ &
 25--$\infty$ & 50--$\infty$ & 100--$\infty$ &
 200--$\infty$ & 500--$\infty$ & 1000--$\infty$\\  \hline
$\sigma_0/\mathrm{pb}$ & $ \!\! 4495.7(2)\!\! $ & $ \!\! 27.589(2)\!\! $ & $ \!\! 1.7906(1)\!\! $ & $ \!\! 0.18128(1)\!\! $ & $ \!\! 0.0065222(4)\!\! $ & $ \!\! 0.00027322(1)\!\! $ \\ \hline
$\de^{\mu^+\nu_\mu}_{\Pq \bar{\Pq}}/\mathrm{\%}$ &                               $  -2.9(1) $ & $  -5.1(1) $ & $  -8.6(1) $ & $  -13.2(1) $ & $  -23.4(1) $ & $  -34.7(1) $ \\                         
$\de^{\mathrm{rec}}_{\Pq \bar{\Pq}}/\mathrm{\%}$ &                               $  -1.8(1) $ & $  -2.6(1) $ & $  -6.1(1) $ & $  -10.3(1) $ & $  -19.5(1) $ & $  -29.5(1) $ \\ \hline                  
$\de_{\Pq \gamma}/\mathrm{\%}$ &                                                 $  0.065(1) $ & $  4.7(1) $ & $  12.3(1) $ & $  17.1(1) $ & $  16.7(1) $ & $  13.5(1) $ \\ \hline                  
$\de^{(2)}_{\mathrm{Sudakov}}/\mathrm{\%}$ &                                     $  -0.0002 $ & $  -0.023 $ & $  -0.082 $ & $  0.057 $ & $  1.3 $ & $  3.8 $ \\ \hline                  
$\de_{\mathrm{multi-} \gamma}/\mathrm{\%}$ &                                     $  0.12^{+ 0.03}_{ -0.02} $ & $  0.31^{+ 0.08}_{ -0.07} $ & $  0.27^{+ 0.06}_{ -0.05} $ & $  0.31^{+ 0.06}_{ -0.06} $ & $  0.41^{+ 0.08}_{ -0.07} $ & $  0.57^{+ 0.10}_{ -0.09} $ \\ \hline                  
$\de^{\mu^+\nu_\mu}_{\mathrm{EW}}/\mathrm{\%}$ &                                 $  -2.7(1) $ & $  0.0(1) $ & $  4.0(1) $ & $  4.3(1) $ & $  -6.3(1) $ & $  -20.6(1) $ \\                         
$\de^{\mathrm{rec}}_{\mathrm{EW}}/\mathrm{\%}$ &                                 $  -1.7(1) $ & $  2.1(1) $ & $  6.2(1) $ & $  6.9(1) $ & $  -2.7(1) $ & $  -16.0(1) $ \\ \hline \hline           
$\de_{\mathrm{QCD}}^{\mu = \MW}/\mathrm{\%}$ &                                   $  -2.7(1) $ & $  812(1) $ & $  784(1) $ & $  814(1) $ & $  611(1) $ & $  399(1) $ \\                         
$\de_{\mathrm{QCD}}^{\mu = M_{\mathrm{T}, \PW}}/\mathrm{\%}$ &                            $  -2.8(1) $ & $  793(1) $ & $  685(1) $ & $  607(1) $ & $  323(1) $ & $  127(1) $ \\ \hline                  
 \end{tabular}

 \ece
\mycaption{\label{ta:pt} Integrated LO cross sections $\sigma_0$
  for \PWp~production at the LHC for different ranges in
  $p_{\mathrm{T},\Pl}$ and corresponding relative corrections $\de$ in
  the SM.}  
\etab

\btab
\bce  \begin{tabular}{c|cccccc}
 \multicolumn{7}{c}{$\Pp\Pp\to \Pl^+ \nu_\Pl \X$ at $\sqrt{s}=14\TeV$}
 \\  \hline$M_{\mathrm{T},\nu_\Pl \Pl}/\mathrm{GeV}$ &
 50--$\infty$ & 100--$\infty$ & 200--$\infty$ &
 500--$\infty$ & 1000--$\infty$ & 2000--$\infty$\\  \hline
$\sigma_0/\mathrm{pb}$ & $ \!\! 4495.7(2)\!\! $ & $ \!\! 27.589(2)\!\! $ & $ \!\! 1.7906(1)\!\! $ & $ \!\! 0.084697(4)\!\! $ & $ \!\! 0.0065222(4)\!\! $ & $ \!\! 0.00027322(1)\!\! $ \\ \hline
$\de^{\mu^+\nu_\mu}_{\Pq \bar{\Pq}}/\mathrm{\%}$ &                               $  -2.9(1) $ & $  -5.2(1) $ & $  -8.1(1) $ & $  -14.8(1) $ & $  -22.6(1) $ & $  -33.2(1) $ \\                         
$\de^{\mathrm{rec}}_{\Pq \bar{\Pq}}/\mathrm{\%}$ &                               $  -1.8(1) $ & $  -3.5(1) $ & $  -6.5(1) $ & $  -12.7(1) $ & $  -20.0(1) $ & $  -29.6(1) $ \\ \hline                  
$\de_{\Pq \gamma}/\mathrm{\%}$ &                                                 $  0.052(1) $ & $  0.12(1) $ & $  0.25(1) $ & $  0.37(1) $ & $  0.39(1) $ & $  0.36(1) $ \\ \hline                  
$\de^{(2)}_{\mathrm{Sudakov}}/\mathrm{\%}$ &                                     $  -0.0002 $ & $  -0.023 $ & $  -0.082 $ & $  0.21 $ & $  1.3 $ & $  3.8 $ \\ \hline                  
$\de_{\mathrm{multi-} \gamma}/\mathrm{\%}$ &                                     $  0.12^{+ 0.03}_{ -0.02} $ & $  0.20^{+ 0.05}_{ -0.04} $ & $  0.16^{+ 0.03}_{ -0.03} $ & $  0.19^{+ 0.04}_{ -0.03} $ & $  0.24^{+ 0.04}_{ -0.04} $ & $  0.34^{+ 0.06}_{ -0.05} $ \\ \hline                  
$\de^{\mu^+\nu_\mu}_{\mathrm{EW}}/\mathrm{\%}$ &                                 $  -2.7(1) $ & $  -4.9(1) $ & $  -7.7(1) $ & $  -14.2(1) $ & $  -22.0(1) $ & $  -32.5(1) $ \\                         
$\de^{\mathrm{rec}}_{\mathrm{EW}}/\mathrm{\%}$ &                                 $  -1.7(1) $ & $  -3.4(1) $ & $  -6.3(1) $ & $  -12.3(1) $ & $  -19.6(1) $ & $  -29.3(1) $ \\ \hline \hline           
$\de_{\mathrm{QCD}}^{\mu = \MW}/\mathrm{\%}$ &                                   $  -4.2(1) $ & $  23.2(1) $ & $  26.6(1) $ & $  19.1(1) $ & $  4.7(1) $ & $  -18.5(1) $ \\                         
$\de_{\mathrm{QCD}}^{\mu = M_{\mathrm{T}, \PW}}/\mathrm{\%}$ &                            $  -4.4(1) $ & $  22.5(1) $ & $  24.0(1) $ & $  12.6(1) $ & $  -6.2(1) $ & $  -34.6(1) $ \\ \hline                  
 \end{tabular}

 \ece
\mycaption{\label{ta:mt} Integrated LO cross sections $\sigma_0$
  for \PWp~production at the LHC for different ranges in $M_{\mathrm{T},
    \nu_\Pl \Pl}$ and corresponding relative corrections $\de$ in the
  SM.}  
\etab

For reference, we first update the ${\cal O}(\alpha)$ NLO corrections
for \PW-boson hadroproduction through the parton process
$\Pq\bar{\Pq}' \to \PW^+ \to \Pl^+ \nu_{\Pl}$
\cite{Dittmaier:2001ay}. The results are given for bare muon final
states ($\de^{\mu^+\nu_\mu}_{\Pq \bar{\Pq}}$) and with photon
recombination applied ($\de^{\mathrm{rec}}_{\Pq \bar{\Pq}}$). As
explained above, the mass-singular corrections $\propto
\alpha\ln(m_\mu/\MW)$ present 
in $\de^{\mu^+\nu_\mu}_{\Pq \bar{\Pq}}$ cancel
if the photon is recombined, rendering the corresponding correction
$\de^{\mathrm{rec}}_{\Pq \bar{\Pq}}$ smaller.  At large
$p_{\mathrm{T},\Pl}$ and $M_{\mathrm{T},\nu_\Pl \Pl}$ the electroweak
corrections are dominated by the ${\cal O}(\alpha)$ Sudakov logarithms
discussed in \refse{se:sudakov}.  Note that the relative ${\cal
  O}(\alpha)$ corrections $\de_{\Pq \bar{\Pq}}$ presented in
Tables~\ref{ta:pt} and \ref{ta:mt} are not very sensitive to the
choice of the PDF and the choice of the $\eta_\Pl$ cut and thus agree
very well with our previous numerical results presented in
\citere{Dittmaier:2001ay}.

\looseness-1
The ${\cal O}(\alpha)$ corrections originating from the photon-induced
processes (\refse{se:Photon-induced}) are not included in $\de_{\Pq
  \bar{\Pq}}$, but are shown separately as $\de_{\gamma \Pq}$ in
Tables~\ref{ta:pt} and \ref{ta:mt}. They are enhanced at large
$p_{\mathrm{T},\Pl}$ because of a new type of contribution where the
incoming photon couples to a \PW~boson that is exchanged in the
$t$-channel.  The photon-induced processes could in principle be used
to extract information on the photon content of the proton. However,
they are overwhelmed by QCD corrections and QCD uncertainties which
strongly affect the $p_{\mathrm{T},\Pl}$ spectrum (see the discussion
of NLO QCD corrections below). If, on the other hand, one considers
the distribution in the transverse mass $M_{\mathrm{T},\nu_{\Pl}
  \Pl}$, which is much less sensitive to QCD effects, the impact of
$\de_{\gamma \Pq}$ is below the percent level. The results for
$\de_{\gamma \Pq}$ in Tables~\ref{ta:pt} and \ref{ta:mt} are in good
agreement with those presented in \citeres{DK_LH, Arbuzov:2007kp}.
Note that in \citere{Arbuzov:2007kp} the mass factorization of the
collinear $\gamma\to \Pq\bar{\Pq}$ splitting is performed in the
\MSbar\ scheme, while the photon distribution of the MRSTQED2004 PDF
set is defined in the DIS scheme as argued above.  It turns out,
however, that the difference between the calculations in the \MSbar\ 
and in the DIS scheme is numerically negligible for hadronic cross
sections.

We find that the ${\cal O}(\alpha^2)$ high-energy Sudakov logarithms
calculated in \refse{se:sudakov}, labeled
$\de^{(2)}_{\mathrm{Sudakov}}$ in Tables~\ref{ta:pt} and \ref{ta:mt},
have a small impact on the cross-section prediction, below 5\% even
for a transverse lepton momentum $p_{\mathrm{T},\Pl}$ in the TeV range.

The corrections due to multi-photon final-state radiation beyond
${\cal O}(\alpha)$ (see Section~\ref{se:Multi-photon}) are shown as
$\de_{\mathrm{multi}-\gamma}$ in the tables.  Only the genuine
higher-order photon effects are included in $\de_{\mathrm{multi}-\gamma}$,
i.e.\ the one-loop contribution is subtracted. We show
$\de_{\mathrm{multi}-\gamma}$ for the central scale choice $Q=
\sqrt{\hat s}$ with an uncertainty estimate obtained from varying the
scale $Q$ between $Q=3 \sqrt{\hat s}$ (upper number) and 
$Q=\sqrt{\hat s}/3$ (lower number).  
Multi-photon final-state radiation beyond ${\cal O}(\alpha)$ has
a very small effect on the cross sections displayed in
Tables~\ref{ta:pt} and \ref{ta:mt}.  The largest part of this small
contribution is in fact due to the change of the coupling constant
$\alpha$ from $\alpha_{\GF}$ to $\alpha(0)$ in the relative ${\cal
  O}(\alpha)$ correction (see also \refse{se:mFSR_comparison}). 
However, as we shall discuss below, the
contribution from multi-photon final-state radiation reaches the
percent level near the \PW~resonance and has thus a significant
impact on a precision determination of the \PW\ mass. A more detailed
analysis is needed to quantify the corresponding shift in the
determination of $\MW$ (cf.~\citeres{Abazov:2003sv, Gerber:2007xk,
  CarloniCalame:2003ux}).

For convenience we combine the above results and display our best
estimate for i) the electroweak corrections for muon final states
$\de^{\mu^+\nu_\mu}_{\mathrm{EW}}$, which includes the ${\cal
  O}(\alpha)$ correction to the $\Pq\bar{\Pq}'$ initial states, the
corrections due to the photon-induced processes, and the multi-photon
final-state radiation corrections with the scale choice $Q=\sqrt{\hat
  s}$, and ii) the total electroweak correction for final states with
photon recombination $\de^{\mathrm{rec}}_{\mathrm{EW}}$, which
combines the ${\cal O}(\alpha)$ corrections from $\Pq\bar{\Pq}'$ and
$\gamma \Pq$ initial states and which is not sensitive to multi-photon
final-state radiation. Because of the theoretical uncertainty in
evaluating the higher-order weak corrections in the high-energy 
regime (see the discussion in \refse{se:sudakov}), we do not include the 
leading two-loop Sudakov logarithms $\de^{(2)}_{\mathrm{Sudakov}}$ in 
our best estimate of the electroweak corrections.

For comparison, we have also calculated the NLO QCD corrections,
evaluated with two different choices for the renormalization and
factorization scales, $\mu_R=\mu_F=\MW$ ($\de_{\mathrm{QCD}}^{\mu =
  \MW}$) and $\mu_R=\mu_F=M_{\mathrm{T}, \PW}$ with $M^2_{\mathrm{T},
  \PW} = \MW^2 + p^2_{\mathrm{T}, \PW}$ ($\de_{\mathrm{QCD}}^{\mu =
  M_{\mathrm{T}, \PW}}$), where $p_{\mathrm{T}, \PW}$ is the
transverse momentum of the \PW~boson to be evaluated on an
event-by-event basis. We have compared our NLO QCD results for
$\de_{\mathrm{QCD}}^{\mu = \MW}$ with those obtained from
MCFM~\cite{mcfm} and find good agreement. In the QCD case, there is of
course only initial-state radiation and thus no technical need for a
recombination procedure of the charged lepton with a possible
additional jet. Rather, one should employ a separation cut
between lepton and jet to allow for a clean event selection.  However,
since we only want to give a rough estimate of the size of QCD effects
for comparison with the electroweak corrections, for simplicity, we do
not include any sort of separation cut. As indicated above, the QCD
corrections are extremely large at large $p_{\mathrm{T},\Pl}$ so that
the electroweak corrections to the $p_{\mathrm{T},\Pl}$ distribution
are overwhelmed by QCD uncertainties.  The $M_{\mathrm{T},\nu_\Pl \Pl}$
distribution, on the other hand, is much less sensitive to
higher-order QCD effects.  It is invariant under transverse boosts to
first order in the velocity of the \PW~boson and thus not strongly
affected by a transverse momentum of the \PW~boson induced by gluon
radiation at NLO QCD~\cite{Smith:1983aa}. 

The size of the QCD corrections to the $p_{\mathrm{T},\Pl}$
distribution can be reduced by applying a jet veto.
Table~\ref{ta:veto} shows the impact of a jet veto on the NLO QCD
calculation where we restrict the additional parton to a transverse
momentum $p_{\mathrm{T}, \mathrm{jet}}~(= p_{\mathrm{T},
  \PW}~\mathrm{at~NLO}) < 50$~GeV. Of course, the jet veto also reduces
the size of the photon-induced processes as demonstrated in
Table~\ref{ta:veto}.

\btab \bce  \begin{tabular}{c|cccccc}
 \multicolumn{7}{c}{$\Pp\Pp\to \Pl^+ \nu_\Pl \X$ at $\sqrt{s}=14\TeV$}
 \\  \hline$p_{\mathrm{T},\Pl}/\mathrm{GeV}$ &
 25--$\infty$ & 50--$\infty$ & 100--$\infty$ &
 200--$\infty$ & 500--$\infty$ & 1000--$\infty$\\  \hline
$\de_{\Pq \gamma, \mathrm{veto}}/\mathrm{\%}$ & $0.025(1)$ & $1.2(1)$ & $0.049(1)$ & $0.043(1)$ & $0.042(1)$ & $0.042(1)$ \\ \hline                  
$\de_{\mathrm{QCD, veto}}^{\mu = \MW}/\mathrm{\%}$ & $  -7.3(1) $ & $  454(1) $ & $  6.4(1) $ & $  -15.8(1) $ & $  -51.9(1) $ & $ -85.0(1) $ \\   \hline                  
 \end{tabular}

 \ece \mycaption{\label{ta:veto}
  Relative corrections for \PWp~production at the LHC from the 
  photon-induced processes and from the NLO QCD calculation with a jet veto
  imposed. We require the additional parton to be produced at
  $p_{\mathrm{T}, \mathrm{jet}} < 50$~GeV.}  \etab

In Figures~\ref{fi:pt} and \ref{fi:mt} we show the differential cross
sections and the corresponding corrections with respect to the
transverse momentum $p_{\mathrm{T},\Pl}$ and the transverse mass 
$M_{\mathrm{T},\nu_\Pl \Pl}$, respectively.  \bfi[fp] \bce
\includegraphics[width=16.cm]{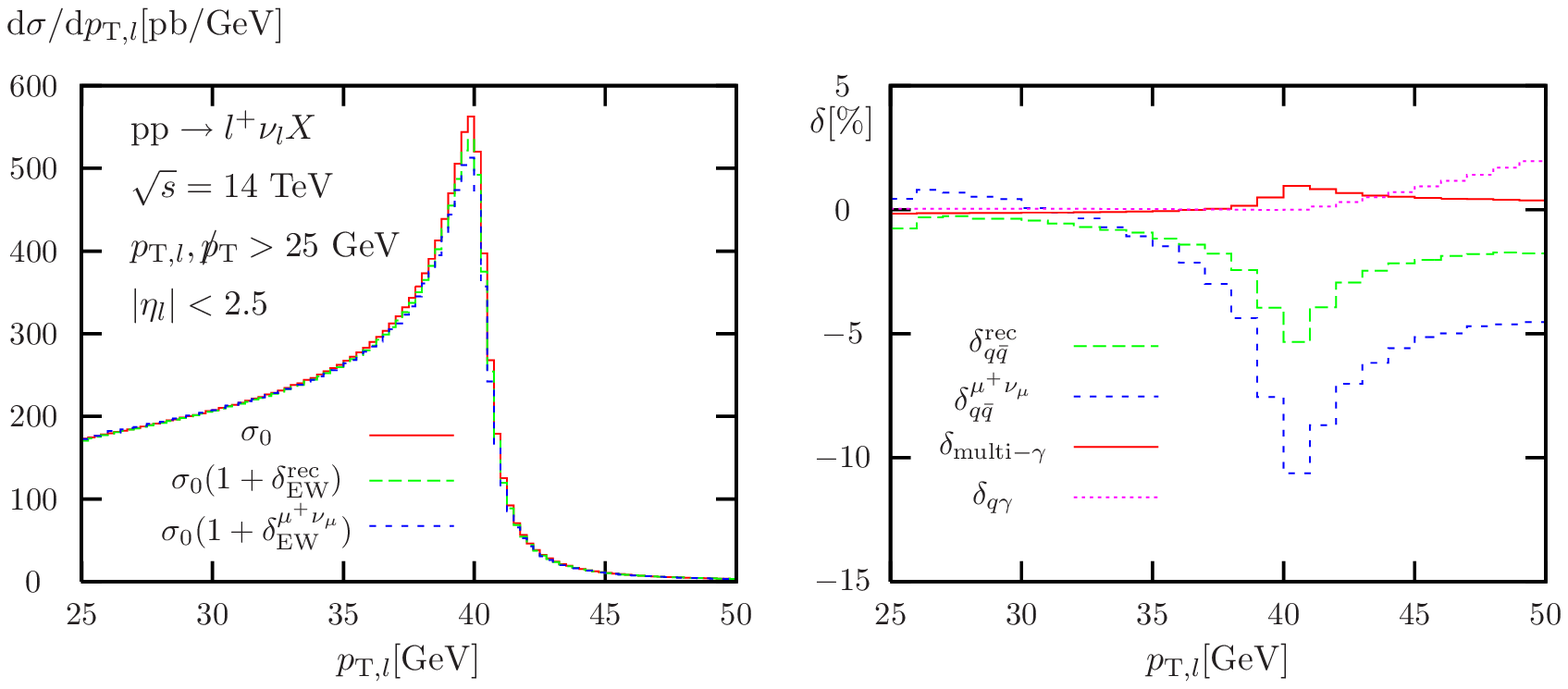} \ece
\mycaption{\label{fi:pt} Lepton-transverse-momentum distribution in LO
  and corresponding relative corrections $\de$ at the LHC in the SM.}
\efi \bfi[fp] \bce
\includegraphics[width=16.cm]{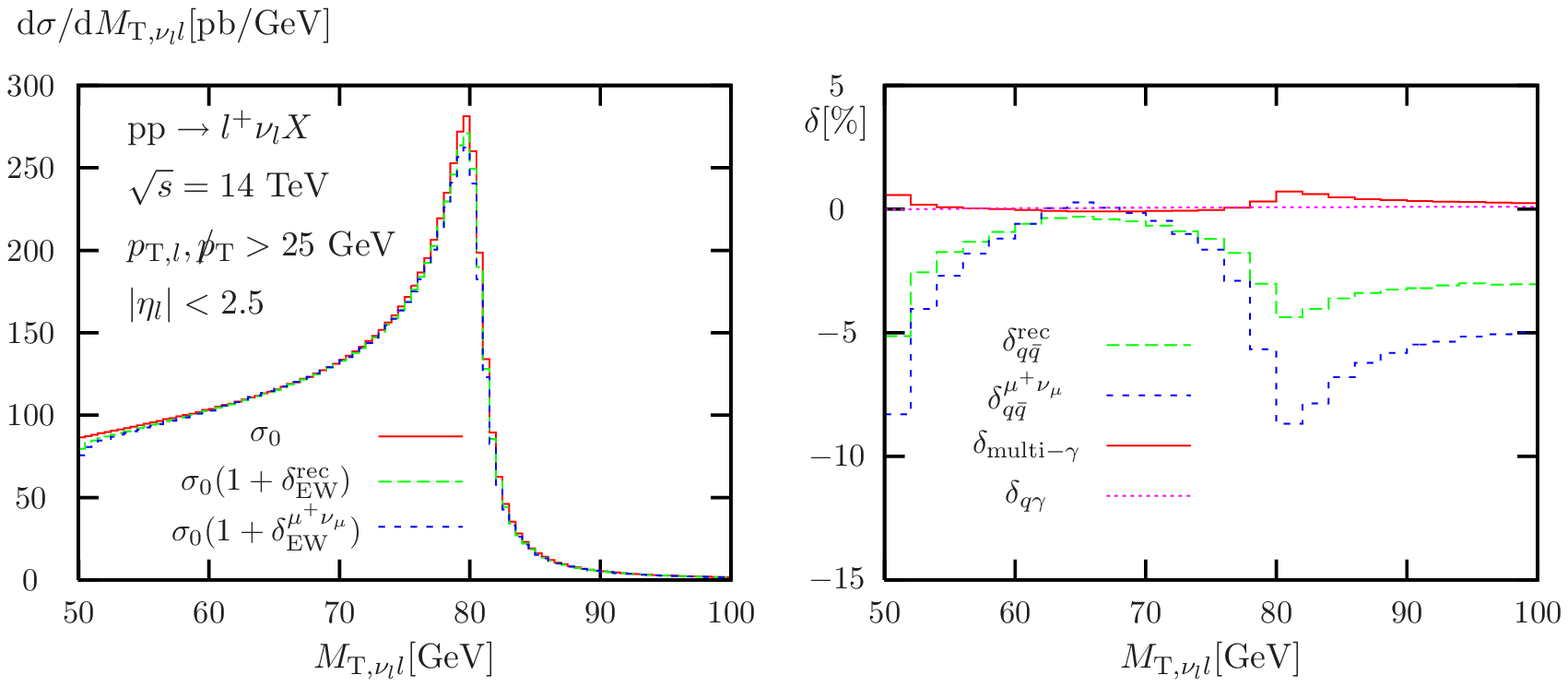} \ece
\mycaption{\label{fi:mt} \PW-transverse-mass distribution in LO and
  corresponding relative corrections $\de$ at the LHC in the SM.}
\efi The distributions show the well-known kinks at
$p_{\mathrm{T},\Pl}\approx \MW/2$ and $M_{\mathrm{T},\nu_\Pl
  \Pl}\approx\MW$, which are used in the \PW-mass determination. Near
these kinks the correction $\de_{\Pq \bar{\Pq}}$ reaches the order of 
10\% for bare muons and is reduced to about 5\% after photon recombination.
Near the resonance region, the corrections from photon-induced
processes are very small. Multi-photon emission, on the other hand,
reaches the percent level near $p_{\mathrm{T},\Pl}\approx \MW/2$ and
$M_{\mathrm{T},\nu_\Pl \Pl}\approx\MW$ and induces some distortion
that affects the $\MW$ determination from the shape of the
$M_{\mathrm{T},\nu_\Pl \Pl}$ distribution. The two-loop Sudakov
corrections are, of course, completely negligible near the \PW\ 
resonance and are thus not displayed in Figures~\ref{fi:pt} and
\ref{fi:mt}.

\subsubsection{\boldmath{Cross sections and distributions for 
    $\Pp\bar{\Pp}\to\PW^+\to \Pl^+ \nu_\Pl \X$ at the
    Tevatron}}

We also present numerical results for \PW~production at the Tevatron,
i.e.\  a $\Pp\bar\Pp$ collider with a CM energy of
$\sqrt{s}=1.96\TeV$.  We again use the phase-space cuts of
\refeq{eq:lcuts} and the photon recombination procedure specified in
the previous section.

Tables~\ref{ta:pt_tev}/\ref{ta:mt_tev} and
Figures~\ref{fi:pt_tev}/\ref{fi:mt_tev} display the LO cross section
$\sigma_0$ and the various relative corrections $\de$ as a function of
the transverse momentum and transverse mass.  The
electroweak corrections are typically of the same size as those
discussed for the LHC, and also show the same qualitative features.
It is obvious that the high-energy Sudakov regime is not
phenomenologically relevant for \PW~production at the Tevatron, and
that the size of the two-loop Sudakov corrections at moderate
$p_{\mathrm{T},\Pl}$ and $M_{\mathrm{T},\nu_\Pl \Pl}$ is no reliable
estimate of the theoretical uncertainty from missing electroweak
two-loop corrections.

\btab[fp]
\bce  \begin{tabular}{c|cccccc}
 \multicolumn{7}{c}{$\Pp\Ppbar\to \Pl^+ \nu_\Pl \X$ at $\sqrt{s}=1.96\TeV$}
 \\  \hline$p_{\mathrm{T},\Pl\nu}/\mathrm{GeV}$ &
 25--$\infty$ & 50--$\infty$ & 75--$\infty$ &
 100--$\infty$ & 200--$\infty$ & 300--$\infty$\\  \hline
$\sigma_0/\mathrm{pb}$ & $ \!\! 706.95(1)\!\! $ & $ \!\! 3.7496(2)\!\! $ & $ \!\! 0.51112(2)\!\! $ & $ \!\! 0.152014(6)\!\! $ & $ \!\! 0.0056405(2)\!\! $ & $ \!\! 0.00039160(2)\!\! $ \\ \hline
$\de^{\mu^+\nu_\mu}_{\Pq \bar{\Pq}}/\mathrm{\%}$ &                               $  -2.7(1) $ & $  -5.4(1) $ & $  -7.4(1) $ & $  -9.1(1) $ & $  -14.2(1) $ & $  -18.6(1) $ \\                         
$\de^{\mathrm{rec}}_{\Pq \bar{\Pq}}/\mathrm{\%}$ &                               $  -1.7(1) $ & $  -2.8(1) $ & $  -4.9(1) $ & $  -6.3(1) $ & $  -10.3(1) $ & $  -13.7(1) $ \\ \hline                  
$\de_{\Pq \gamma}/\mathrm{\%}$ &                                                 $  0.020(1) $ & $  1.5(1) $ & $  2.0(1) $ & $  2.0(1) $ & $  1.4(1) $ & $  0.95(1) $ \\ \hline                  
$\de^{(2)}_{\mathrm{Sudakov}}/\mathrm{\%}$ &                                     $  -0.0001 $ & $  -0.017 $ & $  -0.056 $ & $  -0.078 $ & $  -0.013 $ & $  0.20 $ \\ \hline                  
$\de_{\mathrm{multi-} \gamma}/\mathrm{\%}$ &                                     $  0.11^{+ 0.02}_{ -0.02} $ & $  0.34^{+ 0.09}_{ -0.08} $ & $  0.30^{+ 0.07}_{ -0.06} $ & $  0.32^{+ 0.07}_{ -0.07} $ & $  0.45^{+ 0.09}_{ -0.09} $ & $  0.59^{+ 0.12}_{ -0.11} $ \\ \hline                  
$\de^{\mu^+\nu_\mu}_{\mathrm{EW}}/\mathrm{\%}$ &                                 $  -2.6(1) $ & $  -3.5(1) $ & $  -5.1(1) $ & $  -6.8(1) $ & $  -12.3(1) $ & $  -17.1(1) $ \\                         
$\de^{\mathrm{rec}}_{\mathrm{EW}}/\mathrm{\%}$ &                                 $  -1.6(1) $ & $  -1.3(1) $ & $  -2.9(1) $ & $  -4.3(1) $ & $  -8.9(1) $ & $  -12.8(1) $ \\ \hline \hline           
$\de_{\mathrm{QCD}}^{\mu = \MW}/\mathrm{\%}$ &                                   $  11.2(1) $ & $  377(1) $ & $  205(1) $ & $  174(1) $ & $  113(1) $ & $  74.6(1) $ \\                         
$\de_{\mathrm{QCD}}^{\mu = M_{\mathrm{T}, \PW}}/\mathrm{\%}$ &                            $  11.0(1) $ & $  362(1) $ & $  176(1) $ & $  138(1) $ & $  69.9(1) $ & $  34.9(1) $ \\ \hline                  
 \end{tabular}

 \ece
\mycaption{\label{ta:pt_tev} Integrated LO cross sections $\sigma_0$
  for \PWp~production at the Tevatron for different ranges in
  $p_{\mathrm{T},\Pl}$ and corresponding relative corrections $\de$ in
  the SM.}  
\etab 
\btab[fp] 
\bce  \begin{tabular}{c|cccccc}
 \multicolumn{7}{c}{$\Pp\Ppbar\to \Pl^+ \nu_\Pl \X$ at $\sqrt{s}=1.96\TeV$}
 \\  \hline$M_{\mathrm{T},\nu_\Pl \Pl}/\mathrm{GeV}$ &
 50--$\infty$ & 100--$\infty$ & 150--$\infty$ &
 200--$\infty$ & 400--$\infty$ & 600--$\infty$\\  \hline
$\sigma_0/\mathrm{pb}$ & $ \!\! 706.95(1)\!\! $ & $ \!\! 3.7496(2)\!\! $ & $ \!\! 0.51112(2)\!\! $ & $ \!\! 0.152014(6)\!\! $ & $ \!\! 0.0056405(2)\!\! $ & $ \!\! 0.00039160(2)\!\! $ \\ \hline
$\de^{\mu^+\nu_\mu}_{\Pq \bar{\Pq}}/\mathrm{\%}$ &                               $  -2.7(1) $ & $  -5.2(1) $ & $  -6.5(1) $ & $  -8.0(1) $ & $  -12.7(1) $ & $  -16.8(1) $ \\                         
$\de^{\mathrm{rec}}_{\Pq \bar{\Pq}}/\mathrm{\%}$ &                               $  -1.7(1) $ & $  -3.4(1) $ & $  -4.8(1) $ & $  -6.2(1) $ & $  -10.1(1) $ & $  -13.3(1) $ \\ \hline                  
$\de_{\Pq \gamma}/\mathrm{\%}$ &                                                 $  0.017(1) $ & $  0.028(1) $ & $  0.028(1) $ & $  0.027(1) $ & $  0.018(1) $ & $  0.012(1) $ \\ \hline                  
$\de^{(2)}_{\mathrm{Sudakov}}/\mathrm{\%}$ &                                     $  -0.0001 $ & $  -0.017 $ & $  -0.056 $ & $  -0.078 $ & $  -0.013 $ & $  0.20 $ \\ \hline                  
$\de_{\mathrm{multi-} \gamma}/\mathrm{\%}$ &                                     $  0.11^{+ 0.02}_{ -0.02} $ & $  0.21^{+ 0.05}_{ -0.05} $ & $  0.18^{+ 0.04}_{ -0.04} $ & $  0.19^{+ 0.04}_{ -0.04} $ & $  0.27^{+ 0.05}_{ -0.05} $ & $  0.37^{+ 0.07}_{ -0.07} $ \\ \hline                  
$\de^{\mu^+\nu_\mu}_{\mathrm{EW}}/\mathrm{\%}$ &                                 $  -2.6(1) $ & $  -4.9(1) $ & $  -6.3(1) $ & $  -7.8(1) $ & $  -12.4(1) $ & $  -16.4(1) $ \\                         
$\de^{\mathrm{rec}}_{\mathrm{EW}}/\mathrm{\%}$ &                                 $  -1.6(1) $ & $  -3.4(1) $ & $  -4.8(1) $ & $  -6.1(1) $ & $  -10.0(1) $ & $  -13.3(1) $ \\ \hline \hline           
$\de_{\mathrm{QCD}}^{\mu = \MW}/\mathrm{\%}$ &                                   $  10.8(1) $ & $  22.1(1) $ & $  19.6(1) $ & $  16.7(1) $ & $  6.2(1) $ & $  -2.8(1) $ \\                         
$\de_{\mathrm{QCD}}^{\mu = M_{\mathrm{T}, \PW}}/\mathrm{\%}$ &                            $  10.5(1) $ & $  21.4(1) $ & $  18.5(1) $ & $  15.3(1) $ & $  4.1(1) $ & $  -5.3(1) $ \\ \hline                  
 \end{tabular}

 \ece
\mycaption{\label{ta:mt_tev} Integrated LO cross sections $\sigma_0$
  for \PWp~production at the Tevatron for different ranges in
  $M_{\mathrm{T},\nu_\Pl \Pl}$ and corresponding relative corrections
  $\de$ in the SM.}  \etab 

\bfi[fp] \bce
\includegraphics[width=16.cm]{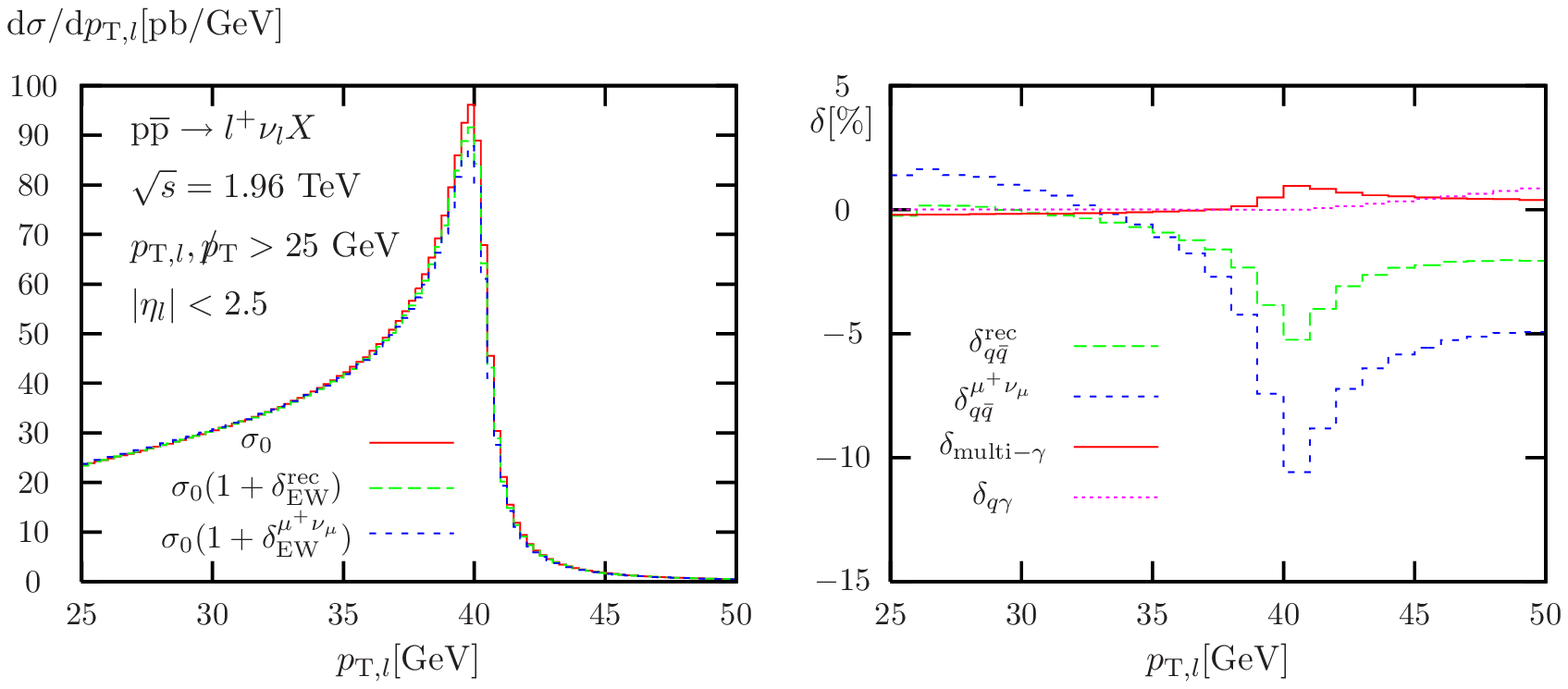} \ece
\mycaption{\label{fi:pt_tev} Lepton-transverse-momentum distribution
  in LO and corresponding relative corrections $\de$ at the Tevatron
  in the SM.}  \efi 
\bfi[fp] \bce
\includegraphics[width=16.cm]{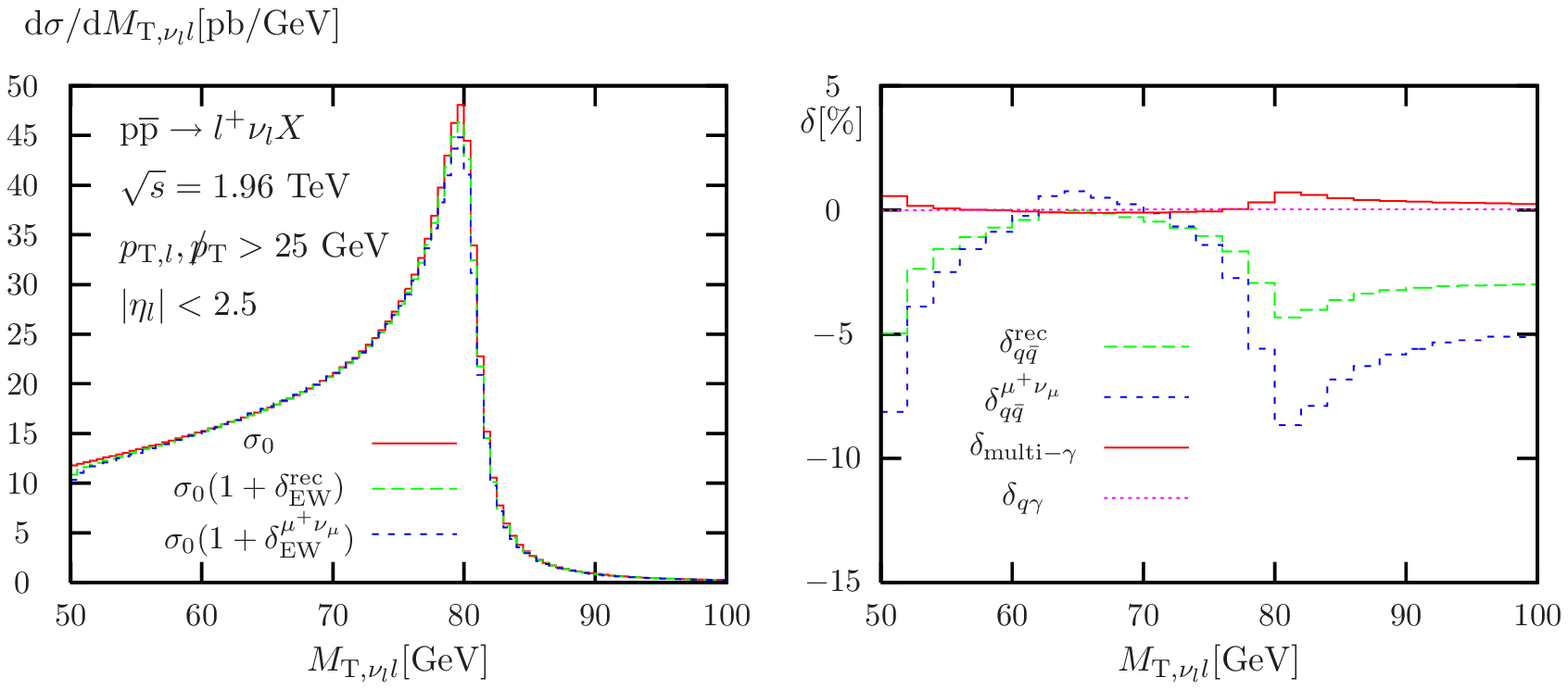} \ece
\mycaption{\label{fi:mt_tev} \PW-transverse-mass distribution in LO and
  corresponding relative corrections $\de$ at the Tevatron in the SM.}
\efi 

\subsubsection{Comparison with existing results on multi-photon 
final-state radiation}
\label{se:mFSR_comparison}

Multi-photon radiation has also been included in the Monte Carlo
programs {\tt Winhac}\cite{Placzek:2003zg} and 
{\tt Horace}\cite{CarloniCalame:2003ux} within a parton-shower approach
in leading logarithmic accuracy.  The program {\tt Horace}, in
particular, combines multi-photon radiation with the
$\mathcal{O}(\alpha)$ electroweak corrections. In the following, we
will compare our results for multi-photon final-state emission within
the structure-function approach defined in
Section~\ref{se:Multi-photon} with the {\tt Horace} results as
presented in~\citere{Gerber:2007xk}.

Adopting the setup of \citere{Gerber:2007xk}, i.e.\ the choice of SM
input parameters, the identification cuts for bare leptons, and the
$\alpha(0)$ input-parameter scheme, we find excellent agreement with
the {\tt Horace} result for the leading order and the ${\cal
  O}(\alpha)$ corrections, as expected from earlier tuned
comparisons~\cite{Buttar:2006zd}.

\begin{figure}
  \begin{center}
  \includegraphics[width=.9\textwidth]{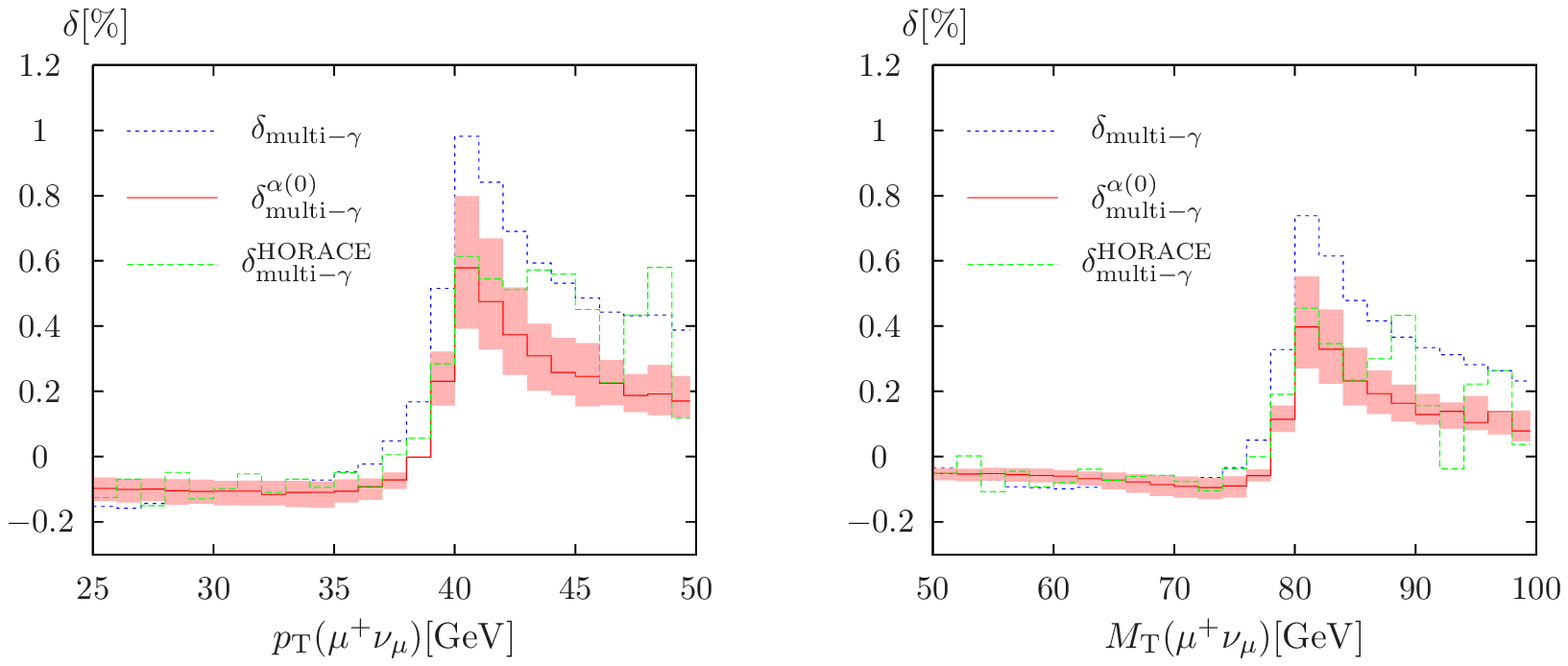}
  \end{center}
  \mycaption{\label{fi:comp_Horace} Comparison of multi-photon
    corrections as obtained in this work with results of {\tt Horace}
    \cite{CarloniCalame:2003ux,Gerber:2007xk} (see text for details).}
\end{figure}

Figure~\ref{fi:comp_Horace} shows the multi-photon final-state
corrections to the distributions in $p_{\mathrm{T},\Pl}$ and
$M_{\mathrm{T},\nu_\Pl \Pl}$ for the muon final state.
Here $\de^{\alpha(0)}_{\mathrm{multi}-\gamma}$ is the result of our
multi-photon correction in the $\alpha(0)$ scheme in the setup of
\citere{Gerber:2007xk}. The band indicates the dependence on the QED
scale \refeq{eq:FSR_scale}, with the factor $\xi$ varied between $1/3$
and $3$.  The result of {\tt Horace} is denoted
$\de^{\mathrm{\tt Horace}}_{\mathrm{multi}-\gamma}$.  For
$p_{\mathrm{T},\Pl}$ and $M_{\mathrm{T},\nu_\Pl \Pl}$ values below and
at the Jacobian peak, the relative corrections differ by less than
$0.1\%$. Above the Jacobian peak the comparison shows larger
deviations (about $0.2\%$) which are comparable to the statistical
fluctuations of the {\tt Horace} results.  The level of agreement is
satisfactory, keeping in mind that the structure-function and
parton-shower approaches involve different approximations. In fact,
the difference between these approximations should be viewed as a lower
bound on the theoretical uncertainty.

Figure~\ref{fi:comp_Horace} also shows the multi-photon correction
$\de_{\mathrm{multi}-\gamma}$, as defined in \refse{se:Multi-photon},
which in contrast to $\de^{\alpha(0)}_{\mathrm{multi}-\gamma}$
contains the formal two-loop effect $(\alpha(0)-\alpha_{\GF})
\ln(\Ml^2/Q^2)$ induced by changing $\alpha_{\GF}$ to $\alpha(0)$ in
the leading photonic ${\cal O}(\alpha)$ correction in the $\GF$
scheme.  Of course, this contribution does not reflect a genuine new
multi-photon effect but only a different partitioning of the total
correction into an $\mathcal{O}(\alpha)$ and a higher-order QED part.

\section{Supersymmetric corrections in the MSSM}
\label{se:mssm}

Non-standard physics could affect the \PW-boson cross-section
prediction and thus bias the precision determination of Standard Model
parameters from \PW-boson observables at hadron colliders. To quantify
the impact of new physics on the \PW~cross section in a concrete
model, we have calculated the ${\cal O}(\alpha)$ electroweak and
${\cal O}(\alpha_{\rm s})$ strong corrections to
$\Pp\Pp/\Pp\bar\Pp\to\PW^+\to \Pl^+ \nu_\Pl \X$ within the MSSM. 

\subsection{Supersymmetric QCD and electroweak corrections}

The SUSY-QCD corrections comprise gluino--squark loops which
contribute to the $\Pq\bar \Pq'\PW$ vertex correction and to the quark
wave-function renormalization.

\begin{figure}
  \begin{center}
  \includegraphics{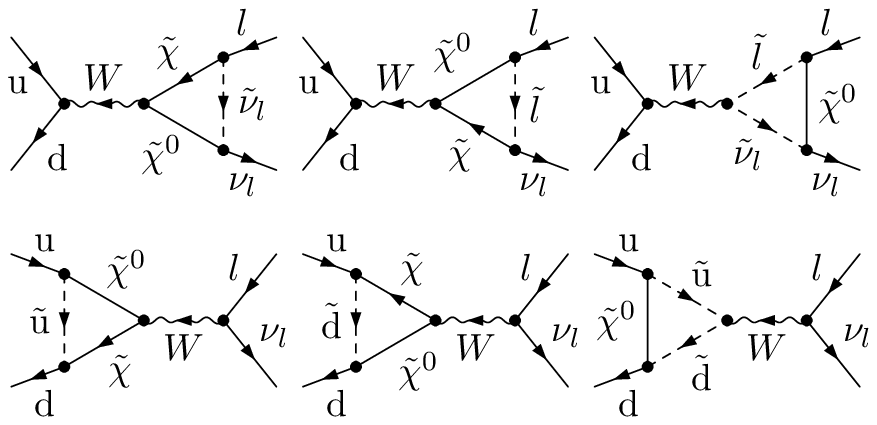}
  \end{center}
  \vspace*{-0.5em} \mycaption{\label{fi:susy_verts} Classes of diagrams
    for the additional vertex corrections within the MSSM. The
    neutralinos $\tilde \chi^0$, charginos $\tilde \chi$, 
    squarks $\tilde \Pq$ ($\Pq=\Pu,\Pd$) and sleptons $\tilde \Pl$
    represent the different possible mass eigenstates.}
\end{figure}

\begin{figure}
  \begin{center}
  \includegraphics{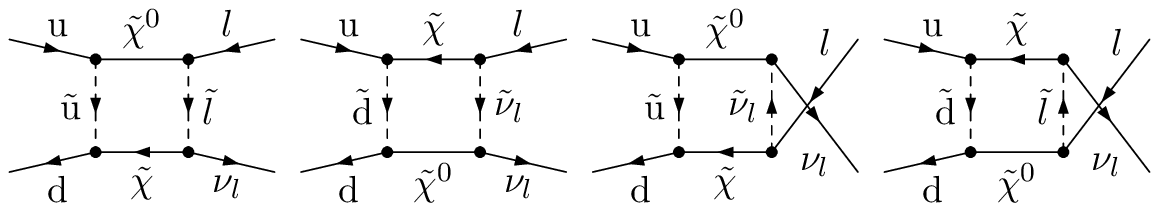}
  \end{center}
  \vspace*{-0.5em} \mycaption{\label{fi:susy_boxes} Classes of diagrams
    for the additional box corrections within the MSSM. The
    neutralinos $\tilde \chi^0$, charginos $\tilde \chi$, 
    squarks $\tilde \Pq$ ($\Pq=\Pu,\Pd$) and sleptons $\tilde \Pl$ 
    represent the different possible mass eigenstates.}
\end{figure}

To obtain the electroweak SUSY corrections, we calculate the complete
electroweak ${\cal O}(\alpha)$ corrections in the MSSM and subtract
the SM corrections, so
that the MSSM corrections can be added to the
SM predictions of the previous section without double counting.  This
procedure applies to the vertex, box, and self-energy corrections as 
well as to the counterterms and $\Delta r$. The diagrams for the 
genuine SUSY vertex and box corrections are shown in 
Figures~\ref{fi:susy_verts} and
\ref{fi:susy_boxes}, respectively. The subtraction is only non-trivial
for contributions from the Higgs sector where the SM and MSSM vertex
appears with different couplings. However, for massless external
fermions the Higgs sector only contributes to the \PW-boson
self-energy and to renormalization constants. The mass of
the lightest MSSM Higgs boson is used as SM input when we subtract the
SM contribution to the correction. Hence, for ultimate precision, one
would have to calculate the SM corrections with the appropriate Higgs
mass.

The diagrammatic calculation is performed in two almost independent
ways. Both calculations use {\tt FeynArts}~\cite{Hahn:2000kx} to
generate the relevant (MSSM) diagrams. One of the calculations then
employs {\tt FormCalc} and {\tt LoopTools}~\cite{Hahn:1998yk} to
perform the algebraic calculation and the loop integrals while the
other calculation uses a completely independent set of in-house
routines for both the algebraic calculation and for the numerical
evaluation.

\subsection{Numerical results}

\subsubsection{Input parameters and setup}

The SM input parameters and the setup of the calculation (input
parameter scheme, PDFs, cuts, etc.) are chosen as described in
\refse{se:SMinput}.

To study the dependence of the corrections on the SUSY breaking
parameters, we show results for all the SPS benchmark
scenarios\cite{Allanach:2002nj}. 
The generic suppression of the genuine SUSY corrections turns out 
to be rather insensitive to a specific scenario. 
We therefore refrain from further restricting the SPS coverage
by taking into account recent experimental bounds in favour of
a broader scope in the SUSY parameter space.
The SPS points are defined by the low-energy SUSY
breaking parameters which determine the spectrum and the couplings.
For the ten benchmark scenarios under consideration in this work, this
input~\cite{SPShomepage} is tabulated in Appendix~\ref{app:SPS}.

Dependent SUSY parameters, such as Higgs, chargino, neutralino, or
sfermion masses, are calculated from the SPS input using tree-level
relations. Since the impact of the fermion masses of the first two
generations is negligible, these masses are set to zero in the
calculation of the corresponding sfermion mass matrices.  Following
this approach the SUSY corrections do not depend on the fermion
generations in the partonic process $\Pu\bar \Pd\to \Pl^+ \nu_\Pl$. 
In particular, the SUSY corrections presented below are valid for both
outgoing electrons and muons.

\subsubsection{Corrections to partonic cross sections}

In order to exhibit the typical features of the SUSY corrections it is
instructive to first discuss the inclusive partonic cross section
$\hat\sigma_{0}$ evaluated at a fixed partonic CM energy
$\sqrt{\hat{s}}$.  Here, no cuts are applied. In
Figure~\ref{fi:partonic} we display the LO result and the SUSY-EW and
SUSY-QCD corrections as a function of the parton energy
$\sqrt{\hat{s}}$ for the ten different SPS scenarios.

\bfi[fp]
\centerline{\setlength{\unitlength}{1cm}
\begin{picture}(13.0,20.9)
\put(0,0){\includegraphics{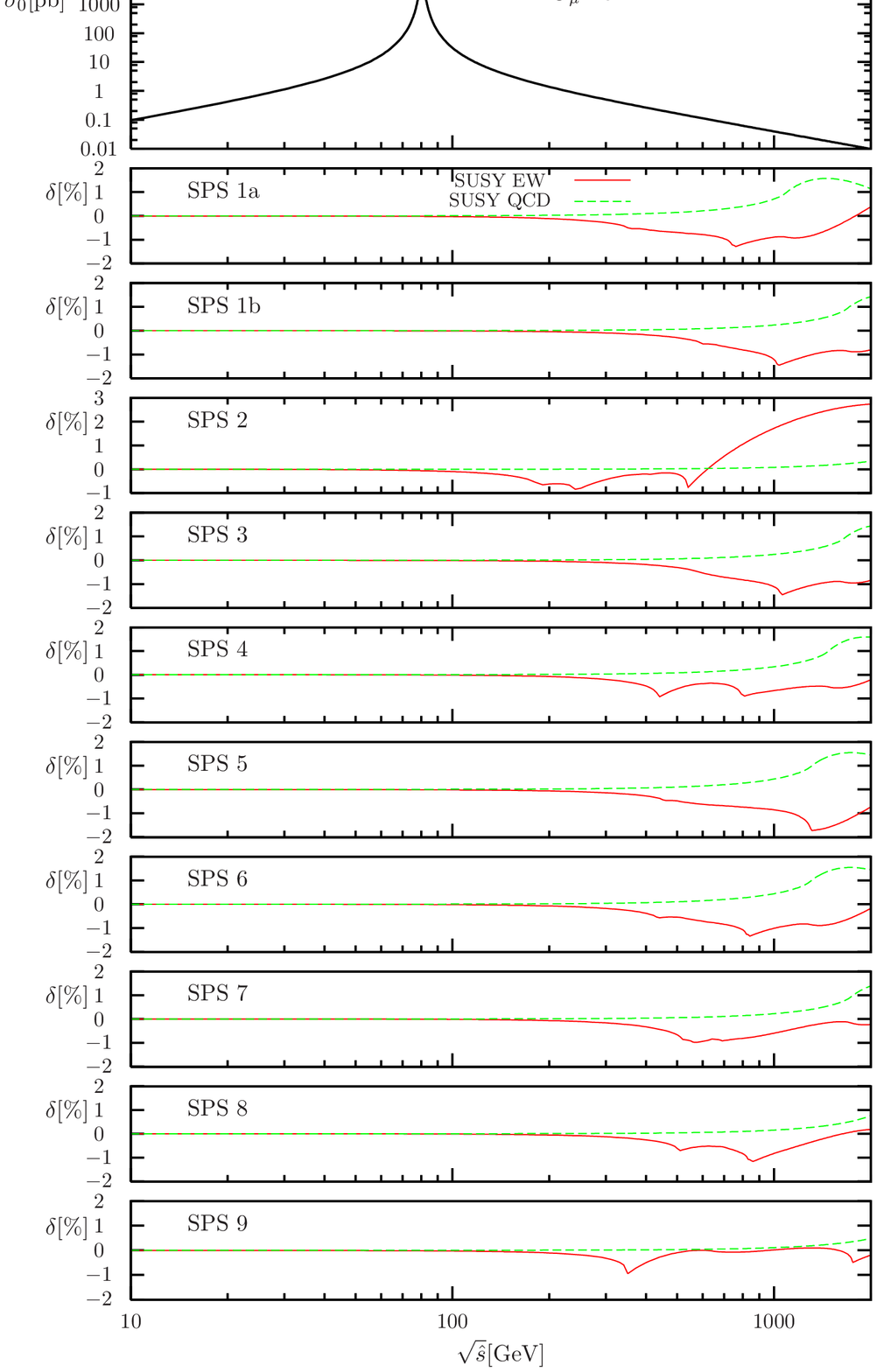}}
\end{picture}
}
\mycaption{\label{fi:partonic}
  Total partonic LO cross section $\hat\sigma_{0}$ and corresponding
  relative SUSY-EW and SUSY-QCD corrections $\de$ as function of the
  partonic CM energy $\sqrt{\hat s}$ for the different SPS scenarios.}
\label{fig:partcs_susy}
\efi

The SUSY-QCD corrections turn out to be completely negligible for
parton energies below $1\TeV$, as the SPS scenarios involve heavy
squarks and gluinos which effectively decouple. The SUSY-QCD
corrections reach the $+1\%$ level only when the sum of the gluino and
a squark mass equals $\sqrt{\hat{s}}$, which typically happens between $1$
and $2\TeV$ for the SPS scenarios. For even larger parton energies
well above the squark/gluino mass scale the SUSY-QCD corrections turn
negative and become logarithmically enhanced with increasing
$\hat{s}$.  In Figure~\ref{fi:partonic}, the corrections are only
shown up to the phenomenologically relevant region $\sqrt{\hat{s}} =
2$ TeV such that the asymptotic behavior is not visible.

The EW corrections turn on at smaller $\sqrt{\hat s}$ and exhibit
peaks which correspond to sparticle thresholds. Specifically,
corrections at the $-1\%$ level occur if the sum of a neutralino and a
chargino mass equals $\sqrt{\hat{s}}$ and if the sleptons and squarks
in the model are heavy compared to this mass scale. The corrections
then rise logarithmically with $\hat{s}$ to positive values. At the
mass scale of the sleptons and squarks there is an additional negative
contribution to the correction which is less peaked than the gaugino
peaks and which washes out the gaugino peaks if the corresponding
regions overlap.

\subsubsection{Corrections to hadronic cross sections}

To calculate the hadronic MSSM cross section we apply the same
phase-space cuts as for the SM analysis. However, since the MSSM
corrections are purely virtual, the kinematics of the relevant events
is as in leading order, i.e.\ photon recombination is irrelevant and 
$M_{\mathrm{T},\nu_\Pl \Pl} = 2 p_{\mathrm{T},\Pl}$.
In Table~\ref{ta:pt_SUSY}, we show the integrated cross section for 
different ranges in $p_{\mathrm{T},\Pl}$ in analogy to the SM analysis in
Section~\ref{se:numres}. As expected, the corrections for relatively
low $p_{\mathrm{T},\Pl}$ cuts are negligible at the sub-per-mille
level.  Only in the high-$p_{\mathrm{T},\Pl}$ tail, the EW corrections
reach the percent level if the SUSY spectrum is light enough.  This
rise in the corrections in the high-$p_{\mathrm{T},\Pl}$ tail depends
on the mass scale of the relevant SUSY particles in the loops. The
maximum of the corrections is reached for the SPS~2 scenario where the
gauginos are particularly light and the squarks and sleptons are so
heavy that their negative contribution becomes effective only at even
larger $p_{\mathrm{T},\Pl}$.

SPS~2 is also the only scenario for which the EW corrections in the
$p_{\mathrm{T},\Pl}$ distribution, as shown in
Figure~\ref{fi:pt_SUSY}, almost reach the percent level for
$p_{\mathrm{T},\Pl} < 100$~GeV due to the light gauginos. In the
\PW-resonance region the corrections are extremely small and flat in
$p_{\mathrm{T},\Pl}$. Only extremely light gauginos with masses
smaller than $\MW$ could impact the precise determination of the $\PW$
mass. Experimental bounds on the mass of the lightest chargino
$M_{\tilde{\chi}^{\pm}} \gsim 100$~GeV~\cite{Yao:2006px} rule out such
a scenario. Note that the SUSY-QCD corrections are multiplied by a
factor of 10 in the plot.

The analogous results for the Tevatron are shown in
Table~\ref{ta:pt_SUSY_tev} and Figure~\ref{fi:pt_SUSY_tev}.

\begin{table}
\begin{center}
 \begin{tabular}{llllllll}
 \multicolumn{8}{c}{$\Pp\Pp\to \Pl^+ \nu_\Pl \X$ at $\sqrt{s}=14 \TeV$}
 \\  \hline& $p_{\mathrm{T},l}/\mathrm{GeV}$ &
 25--$\infty$ & 50--$\infty$ & 100--$\infty$ &
 200--$\infty$ & 500--$\infty$ & 1000--$\infty$\\  \hline
SPS1a & $\de_{\SUSY-\QCD}/\%$ &   $   +0.0046 $ & $    +0.017 $ & $    +0.079 $ & $     +0.29 $ & $      +1.3 $ & $     +0.42 $ \\
SPS1a & $\de_{\SUSY-\EW}/\%$  &   $    -0.022 $ & $    -0.077 $ & $     -0.34 $ & $     -0.78 $ & $     -0.54 $ & $      +1.2 $ \\ \hline
SPS1b & $\de_{\SUSY-\QCD}/\%$ &   $   +0.0019 $ & $   +0.0071 $ & $    +0.032 $ & $     +0.11 $ & $     +0.60 $ & $      +1.3 $ \\
SPS1b & $\de_{\SUSY-\EW}/\%$  &   $   -0.0068 $ & $    -0.028 $ & $     -0.14 $ & $     -0.50 $ & $      -1.0 $ & $     -0.20 $ \\ \hline
SPS2  & $\de_{\SUSY-\QCD}/\%$ &   $   +0.0007 $ & $   +0.0026 $ & $    +0.012 $ & $    +0.039 $ & $     +0.19 $ & $     +0.71 $ \\
SPS2  & $\de_{\SUSY-\EW}/\%$  &   $    -0.061 $ & $     -0.24 $ & $     -0.44 $ & $     +0.21 $ & $      +2.3 $ & $      +2.6 $ \\ \hline
SPS3  & $\de_{\SUSY-\QCD}/\%$ &   $   +0.0020 $ & $   +0.0073 $ & $    +0.033 $ & $     +0.12 $ & $     +0.62 $ & $      +1.3 $ \\
SPS3  & $\de_{\SUSY-\EW}/\%$  &   $   -0.0090 $ & $    -0.031 $ & $     -0.15 $ & $     -0.51 $ & $      -1.0 $ & $     -0.23 $ \\ \hline
SPS4  & $\de_{\SUSY-\QCD}/\%$ &   $   +0.0027 $ & $   +0.0097 $ & $    +0.044 $ & $     +0.16 $ & $     +0.83 $ & $      +1.2 $ \\
SPS4  & $\de_{\SUSY-\EW}/\%$  &   $   -0.0095 $ & $    -0.050 $ & $     -0.25 $ & $     -0.57 $ & $     -0.44 $ & $     +0.53 $ \\ \hline
SPS5  & $\de_{\SUSY-\QCD}/\%$ &   $   +0.0033 $ & $    +0.012 $ & $    +0.055 $ & $     +0.20 $ & $     +0.99 $ & $     +0.89 $ \\
SPS5  & $\de_{\SUSY-\EW}/\%$  &   $    -0.016 $ & $    -0.046 $ & $     -0.20 $ & $     -0.60 $ & $      -1.2 $ & $     +0.12 $ \\ \hline
SPS6  & $\de_{\SUSY-\QCD}/\%$ &   $   +0.0033 $ & $    +0.012 $ & $    +0.055 $ & $     +0.20 $ & $     +1.00 $ & $     +0.88 $ \\
SPS6  & $\de_{\SUSY-\EW}/\%$  &   $    -0.013 $ & $    -0.050 $ & $     -0.24 $ & $     -0.70 $ & $     -0.70 $ & $     +0.61 $ \\ \hline
SPS7  & $\de_{\SUSY-\QCD}/\%$ &   $   +0.0019 $ & $   +0.0067 $ & $    +0.030 $ & $     +0.11 $ & $     +0.57 $ & $      +1.3 $ \\
SPS7  & $\de_{\SUSY-\EW}/\%$  &   $   -0.0091 $ & $    -0.045 $ & $     -0.23 $ & $     -0.69 $ & $     -0.27 $ & $     +0.25 $ \\ \hline
SPS8  & $\de_{\SUSY-\QCD}/\%$ &   $   +0.0013 $ & $   +0.0048 $ & $    +0.022 $ & $    +0.075 $ & $     +0.39 $ & $      +1.4 $ \\
SPS8  & $\de_{\SUSY-\EW}/\%$  &   $   -0.0067 $ & $    -0.035 $ & $     -0.18 $ & $     -0.55 $ & $     -0.30 $ & $     +0.23 $ \\ \hline
SPS9  & $\de_{\SUSY-\QCD}/\%$ &   $   +0.0009 $ & $   +0.0034 $ & $    +0.015 $ & $    +0.051 $ & $     +0.26 $ & $     +0.97 $ \\
SPS9  & $\de_{\SUSY-\EW}/\%$  &   $    -0.021 $ & $    -0.070 $ & $     -0.27 $ & $     -0.10 $ & $    +0.015 $ & $    +0.023 $ \\ \hline
 \end{tabular}

\end{center}
\mycaption{\label{ta:pt_SUSY}
Relative SUSY-EW and SUSY-QCD corrections $\delta$ in the MSSM
for \PWp~production at the LHC for different ranges in 
$p_{\mathrm{T},\Pl}$. The corresponding integrated LO cross sections 
$\sigma_0$ can be found in Table~\ref{ta:pt}.}
\end{table}

\bfi[fp]
\centerline{\setlength{\unitlength}{1cm}
\begin{picture}(13.0,21.3)
\put(0,0){\includegraphics{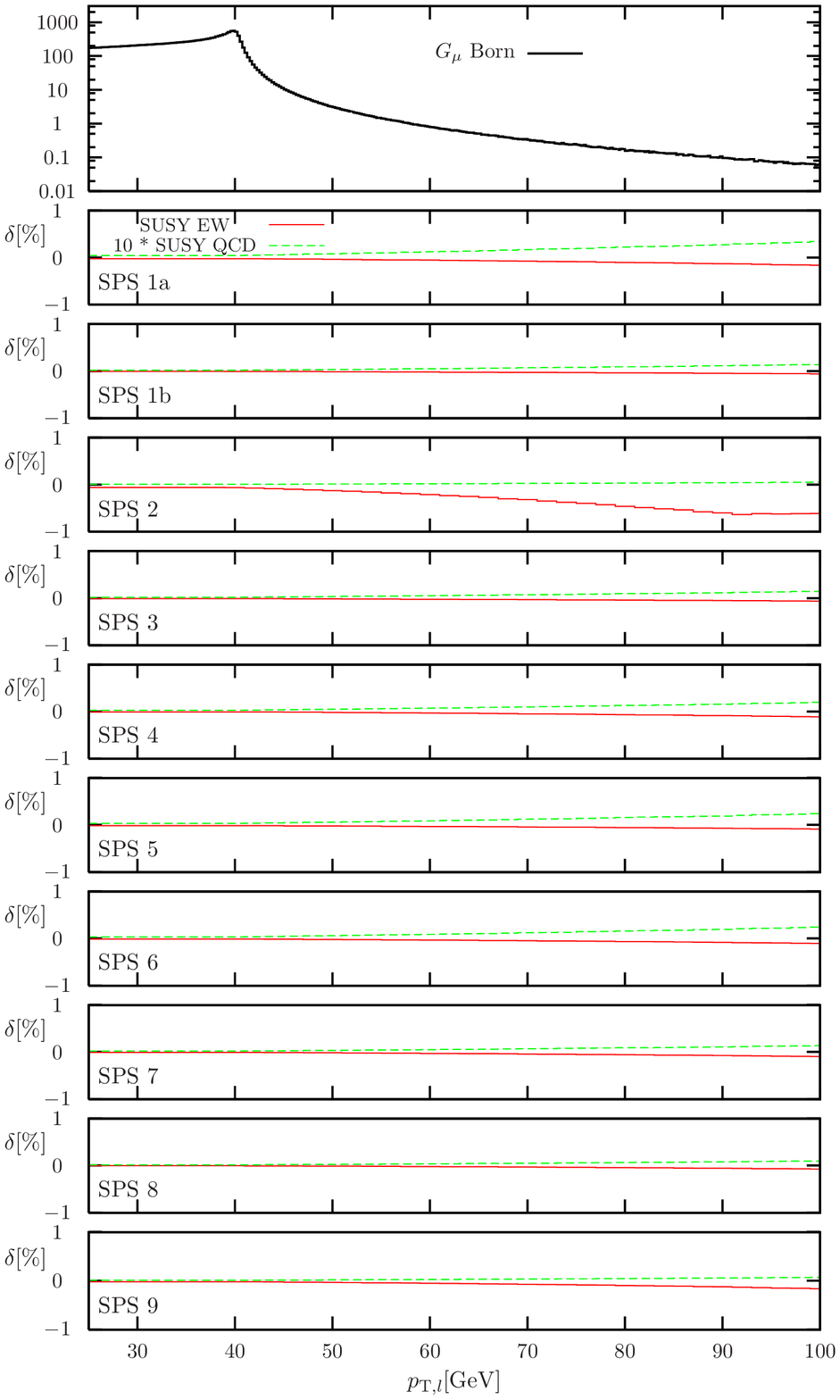}}
\end{picture}
} 
\mycaption{\label{fi:pt_SUSY}
Lepton-transverse-momentum distribution in LO
and corresponding relative SUSY-EW and SUSY-QCD corrections $\de$ 
for the different SPS scenarios at the LHC.}
\efi

\begin{table}
\begin{center}
 \begin{tabular}{llllllll}
 \multicolumn{8}{c}{$\Pp\Ppbar\to \Pl^+ \nu_\Pl \X$ at $\sqrt{s}=1.96\TeV$}
 \\  \hline& $p_{\mathrm{T},l}/\mathrm{GeV}$ &
 25--$\infty$ & 50--$\infty$ & 75--$\infty$ &
 100--$\infty$ & 200--$\infty$ & 300--$\infty$\\  \hline
SPS1a & $\de_{\SUSY-\QCD}/\%$ &   $   +0.0046 $ & $    +0.014 $ & $    +0.032 $ & $    +0.052 $ & $     +0.16 $ & $     +0.32 $ \\
SPS1a & $\de_{\SUSY-\EW}/\%$  &   $    -0.022 $ & $    -0.063 $ & $     -0.16 $ & $     -0.27 $ & $     -0.71 $ & $     -0.97 $ \\ \hline
SPS1b & $\de_{\SUSY-\QCD}/\%$ &   $   +0.0019 $ & $   +0.0057 $ & $    +0.013 $ & $    +0.021 $ & $    +0.064 $ & $     +0.12 $ \\
SPS1b & $\de_{\SUSY-\EW}/\%$  &   $   -0.0068 $ & $    -0.022 $ & $    -0.054 $ & $    -0.093 $ & $     -0.34 $ & $     -0.69 $ \\ \hline
SPS2  & $\de_{\SUSY-\QCD}/\%$ &   $   +0.0007 $ & $   +0.0021 $ & $   +0.0049 $ & $   +0.0079 $ & $    +0.023 $ & $    +0.044 $ \\
SPS2  & $\de_{\SUSY-\EW}/\%$  &   $    -0.061 $ & $     -0.23 $ & $     -0.51 $ & $     -0.54 $ & $     -0.15 $ & $     +0.58 $ \\ \hline
SPS3  & $\de_{\SUSY-\QCD}/\%$ &   $   +0.0020 $ & $   +0.0059 $ & $    +0.014 $ & $    +0.022 $ & $    +0.066 $ & $     +0.13 $ \\
SPS3  & $\de_{\SUSY-\EW}/\%$  &   $   -0.0089 $ & $    -0.024 $ & $    -0.058 $ & $    -0.099 $ & $     -0.35 $ & $     -0.72 $ \\ \hline
SPS4  & $\de_{\SUSY-\QCD}/\%$ &   $   +0.0026 $ & $   +0.0077 $ & $    +0.018 $ & $    +0.029 $ & $    +0.088 $ & $     +0.17 $ \\
SPS4  & $\de_{\SUSY-\EW}/\%$  &   $   -0.0094 $ & $    -0.040 $ & $     -0.11 $ & $     -0.20 $ & $     -0.61 $ & $     -0.48 $ \\ \hline
SPS5  & $\de_{\SUSY-\QCD}/\%$ &   $   +0.0032 $ & $   +0.0095 $ & $    +0.022 $ & $    +0.036 $ & $     +0.11 $ & $     +0.21 $ \\
SPS5  & $\de_{\SUSY-\EW}/\%$  &   $    -0.016 $ & $    -0.037 $ & $    -0.085 $ & $     -0.14 $ & $     -0.49 $ & $     -0.69 $ \\ \hline
SPS6  & $\de_{\SUSY-\QCD}/\%$ &   $   +0.0033 $ & $   +0.0096 $ & $    +0.022 $ & $    +0.037 $ & $     +0.11 $ & $     +0.22 $ \\
SPS6  & $\de_{\SUSY-\EW}/\%$  &   $    -0.012 $ & $    -0.040 $ & $     -0.10 $ & $     -0.18 $ & $     -0.60 $ & $     -0.89 $ \\ \hline
SPS7  & $\de_{\SUSY-\QCD}/\%$ &   $   +0.0018 $ & $   +0.0054 $ & $    +0.013 $ & $    +0.020 $ & $    +0.061 $ & $     +0.12 $ \\
SPS7  & $\de_{\SUSY-\EW}/\%$  &   $   -0.0090 $ & $    -0.035 $ & $    -0.094 $ & $     -0.17 $ & $     -0.65 $ & $     -0.85 $ \\ \hline
SPS8  & $\de_{\SUSY-\QCD}/\%$ &   $   +0.0013 $ & $   +0.0039 $ & $   +0.0090 $ & $    +0.015 $ & $    +0.043 $ & $    +0.081 $ \\
SPS8  & $\de_{\SUSY-\EW}/\%$  &   $   -0.0066 $ & $    -0.027 $ & $    -0.072 $ & $     -0.13 $ & $     -0.48 $ & $     -0.65 $ \\ \hline
SPS9  & $\de_{\SUSY-\QCD}/\%$ &   $   +0.0009 $ & $   +0.0027 $ & $   +0.0063 $ & $    +0.010 $ & $    +0.030 $ & $    +0.057 $ \\
SPS9  & $\de_{\SUSY-\EW}/\%$  &   $    -0.021 $ & $    -0.060 $ & $     -0.15 $ & $     -0.27 $ & $     -0.16 $ & $    -0.048 $ \\ \hline
 \end{tabular}

\end{center}
\mycaption{\label{ta:pt_SUSY_tev}
Relative SUSY-EW and SUSY-QCD corrections $\delta$ in the MSSM
for \PWp~production at the Tevatron for different ranges in 
$p_{\mathrm{T},\Pl}$. The corresponding integrated LO cross sections 
$\sigma_0$ can be found in Table~\ref{ta:pt_tev}.}
\end{table}

\bfi[p]
\centerline{\setlength{\unitlength}{1cm}
\begin{picture}(13.0,21.3)
\put(0,0){\includegraphics{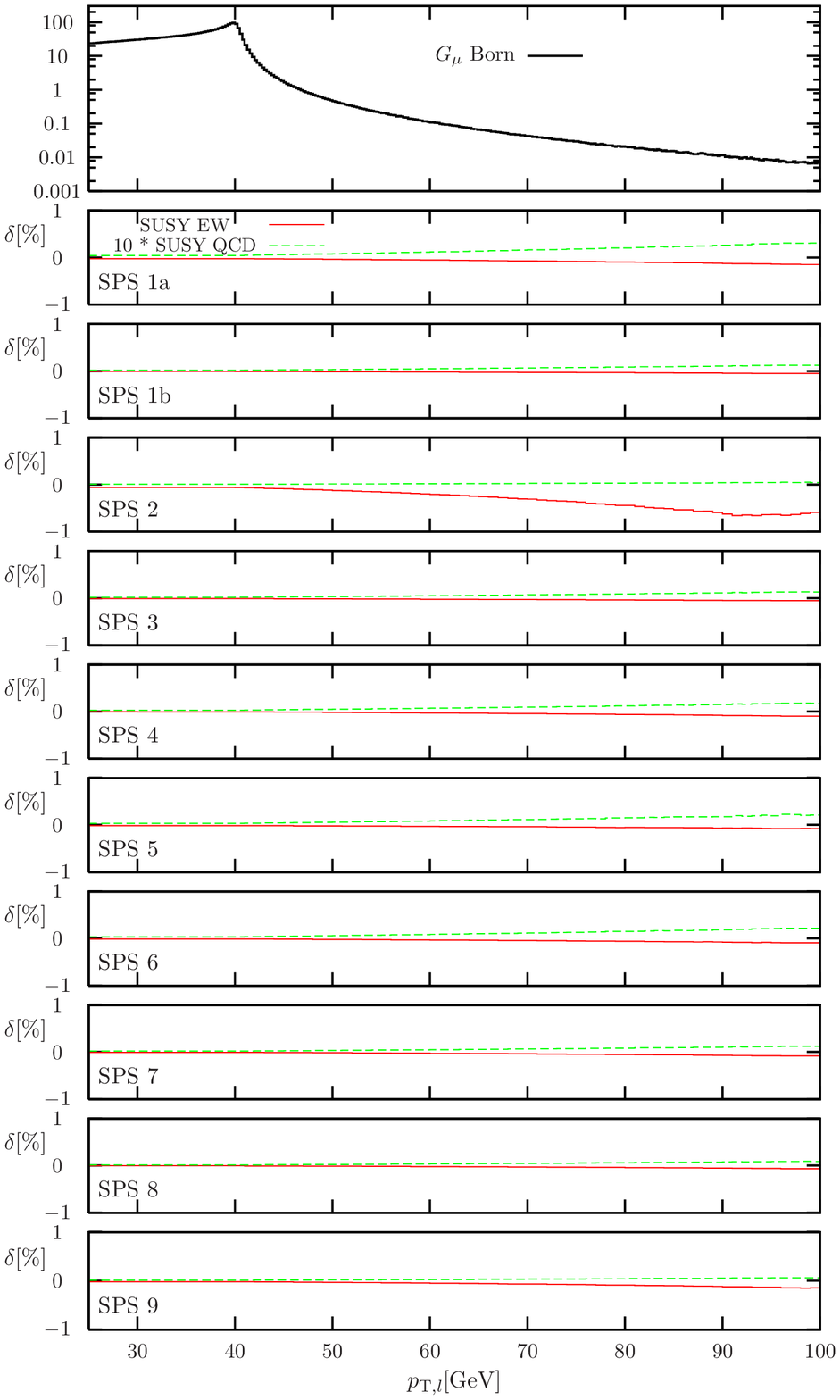}}
\end{picture}
} 
\mycaption{\label{fi:pt_SUSY_tev}
Lepton-transverse-momentum distribution in LO
and corresponding relative SUSY-EW and SUSY-QCD corrections $\de$ 
for the different SPS scenarios at the Tevatron.}
\efi

\section{Conclusions}
\label{se:conclusion}

Single-\PW-boson production is one of the cleanest hadron collider
processes and will be used to precisely determine Standard Model
parameters like the \PW-boson mass and width. In order to match the
envisaged high experimental accuracy, it is mandatory to reduce the
theoretical uncertainty of the cross-section prediction to a level of
one percent or better. 

We have studied radiative corrections to single-\PW-boson production,
$\Pp\Pp/\Pp\bar\Pp \to \PW^+ \to \Pl^+\nu_\Pl \X$, at the LHC and at
the Tevatron. We have completed our previous calculation of the ${\cal
  O}(\alpha)$ electroweak corrections \cite{Dittmaier:2001ay} by
including photon--quark scattering processes.  These photon-induced
contributions turn out to be considerable at large lepton transverse
momentum. However, they do not significantly affect the distribution
in the lepton--neutrino transverse mass $M_{\mathrm{T},\nu_\Pl \Pl}$.
We have furthermore discussed the impact of electroweak effects beyond
${\cal O}(\alpha)$, specifically the Sudakov logarithms which arise in
the high-energy regime. The leading ${\cal O}(\alpha^2)$ Sudakov
logarithms, which we consider as a measure for the electroweak
two-loop effects, turn out to be small, below 5\% even for transverse
lepton momenta $p_{\mathrm{T},\Pl}$ in the TeV range.  Corrections due
to multi-photon final-state radiation beyond ${\cal O}(\alpha)$ reach
the percent level near the \PW~resonance and distort the shape of the
$p_{\mathrm{T},\Pl}$ and $M_{\mathrm{T},\nu_\Pl \Pl}$ distributions.
Given the high experimental accuracy envisaged specifically at the
LHC, it is therefore mandatory to control the effects of multi-photon
emission in the determination of the \PW\ mass.

Finally, to study the impact of new physics on the
\PW\ cross section in a concrete model, we have calculated the ${\cal
  O}(\alpha)$ electroweak and ${\cal O}(\alpha_{\rm s})$ strong
corrections to $\Pp\Pp/\Pp\bar\Pp\to\PW^+\to \Pl^+ \nu_\Pl \X$ within
the MSSM.  The supersymmetric corrections turn out to be negligible in
the vicinity of the W resonance for viable MSSM scenarios, reaching
the percent level only at very high lepton transverse momenta and for
specific choices of the supersymmetric parameters. Effects from
virtual SUSY particles would thus not spoil the status 
of single-\PW-boson production as one of the cleanest Standard Model
candles at hadron colliders.

\section*{Acknowledgments}
We are grateful to Ansgar Denner and Stefano Pozzorini for discussions
about electroweak high-energy logarithms. We would like to thank Carlo
Carloni Calame, Guido Montagna, Oreste Nicrosini and Alessandro Vicini
for providing us with the {\tt Horace} results.  This work was
supported in part by the BMBF grant 05HT6PAA, by the DFG SFB/TR9
"Computational Particle Physics", by the DFG Research Training Group
"Elementary Particle Physics at the TeV Scale", and by the European
Community's Marie-Curie Research Training Network HEPTOOLS under
contract MRTN-CT-2006-035505.  SD and MK thank the Galileo Galilei
Institute for Theoretical Physics in Florence for the hospitality and
the INFN for partial support during the completion of this work.

\begin{appendix}

\section{SPS benchmark scenarios}
\label{app:SPS}

For the SPS benchmark~\cite{Allanach:2002nj} scenarios discussed in
this work we use the low-energy input specified in
Table~\ref{ta:SUSY_input}.  The input variables are the ratio $\tanb$
of the vacuum expectation values of the Higgs bosons giving rise to
up- and down-type fermion masses, the mass of the CP-odd Higgs boson,
$\MA$, the supersymmetric Higgs mass parameter $\mu$, the electroweak
gaugino mass parameters $M_{1,2}$, the gluino mass $m_{\tilde{g}}$,
the trilinear couplings $A_{\tau,\Pt,\Pb}$, the scale at which the
\DRbar-input values are defined, $\mu_R ($\DRbar$)$, the soft
SUSY-breaking parameters in the diagonal entries of the squark and
slepton mass matrices of the first and second generations $M_{fi}$
(where $i=L,R$ refers to the left- and right-handed sfermions, $\Pf=\Pq,\Pl$
to quarks and leptons, and $\Pf=\Pu,\Pd,\Pe$ to up and down quarks and
electrons, respectively), and the analogous soft SUSY-breaking
parameters for the third generation $M^{3G}_{fi}$. 
\btab
\begin{center}
\vspace{18cm}
\begin{rotate}{90}
\begin{tabular}{|r|r|r|r|r|r|r|r|r|r|r|}
\hline
\mbox{} \hspace{1cm} \mbox{} & SPS~1a     & SPS~1b    & SPS~2    & SPS~3         & SPS~4      & SPS~5       & SPS~6    & SPS~7        & SPS~8      & SPS~9   \\ \hline
$\tanb$                      & $     10 $ & $    30 $ & $     10 $ & $      10 $ & $     50 $ & $       5 $ & $     10 $ & $     15 $ & $     15 $ & $     10 $  \\
$\MA$[GeV]                   & $  393.6 $ & $ 525.5 $ & $ 1443.0 $ & $   572.4 $ & $  404.4 $ & $   693.9 $ & $  463.0 $ & $  377.9 $ & $  514.5 $ & $  911.7 $  \\
$\mu$[GeV]                   & $  352.4 $ & $ 495.6 $ & $  124.8 $ & $   508.6 $ & $  377.0 $ & $   639.8 $ & $  393.9 $ & $  300.0 $ & $  398.3 $ & $  869.9 $  \\
$M_1$[GeV]                   & $   99.1 $ & $ 162.8 $ & $  120.4 $ & $   162.8 $ & $  120.8 $ & $   121.4 $ & $  195.9 $ & $  168.6 $ & $  140.0 $ & $ -550.6 $  \\
$M_2$[GeV]                   & $  192.7 $ & $ 310.9 $ & $  234.1 $ & $   311.4 $ & $  233.2 $ & $   234.6 $ & $  232.1 $ & $  326.8 $ & $  271.8 $ & $ -175.5 $  \\
$m_{\tilde{g}}$[GeV]         & $  595.2 $ & $ 916.1 $ & $  784.4 $ & $   914.3 $ & $  721.0 $ & $   710.3 $ & $  708.5 $ & $  926.0 $ & $  820.5 $ & $ 1275.2 $  \\
$A_\tau$[GeV]                & $ -254.2 $ & $-195.8 $ & $ -187.8 $ & $  -246.1 $ & $ -102.3 $ & $ -1179.3 $ & $ -213.4 $ & $  -39.0 $ & $  -36.7 $ & $ 1162.4 $  \\
$A_\Pt$[GeV]                 & $ -510.0 $ & $-729.3 $ & $ -563.7 $ & $  -733.5 $ & $ -552.2 $ & $  -905.6 $ & $ -570.0 $ & $ -319.4 $ & $ -296.7 $ & $ -350.3 $  \\
$A_\Pb$[GeV]                 & $ -772.7 $ & $-987.4 $ & $ -797.2 $ & $ -1042.2 $ & $ -729.5 $ & $ -1671.4 $ & $ -811.3 $ & $ -350.5 $ & $ -330.3 $ & $  216.4 $  \\
$\mu_R ($\DRbar$)$[GeV]      & $  454.7 $ & $ 706.9 $ & $ 1077.1 $ & $   703.8 $ & $  571.3 $ & $   449.8 $ & $  548.3 $ & $  839.6 $ & $  987.8 $ & $ 1076.1 $  \\ 
$M_{\Pq L}$[GeV]             & $  539.9 $ & $ 836.2 $ & $ 1533.6 $ & $   818.3 $ & $  732.2 $ & $   643.9 $ & $  641.3 $ & $  861.3 $ & $ 1081.6 $ & $ 1219.2 $  \\
$M_{\Pd R}$[GeV]             & $  519.5 $ & $ 803.9 $ & $ 1530.3 $ & $   788.9 $ & $  713.9 $ & $   622.9 $ & $  621.8 $ & $  828.6 $ & $ 1029.0 $ & $ 1237.6 $  \\
$M_{\Pu R}$[GeV]             & $  521.7 $ & $ 807.5 $ & $ 1530.5 $ & $   792.6 $ & $  716.0 $ & $   625.4 $ & $  629.3 $ & $  831.3 $ & $ 1033.8 $ & $ 1227.9 $  \\
$M_{\Pl L}$[GeV]             & $  196.6 $ & $ 334.0 $ & $ 1455.6 $ & $   283.3 $ & $  445.9 $ & $   252.2 $ & $  260.7 $ & $  257.2 $ & $  353.5 $ & $  316.2 $  \\
$M_{eR}$[GeV]                & $  136.2 $ & $ 248.3 $ & $ 1451.0 $ & $   173.0 $ & $  414.2 $ & $   186.8 $ & $  232.8 $ & $  119.7 $ & $  170.4 $ & $  300.0 $  \\
$M^{3G}_{\Pq L}$[GeV]        & $  495.9 $ & $ 762.5 $ & $ 1295.3 $ & $   760.7 $ & $  640.1 $ & $   535.2 $ & $  591.2 $ & $  836.3 $ & $ 1042.7 $ & $ 1111.6 $  \\
$M^{3G}_{\Pd R}$[GeV]        & $  516.9 $ & $ 780.3 $ & $ 1519.9 $ & $   785.6 $ & $  673.4 $ & $   620.5 $ & $  619.0 $ & $  826.9 $ & $ 1025.5 $ & $ 1231.7 $  \\
$M^{3G}_{\Pu R}$[GeV]        & $  424.8 $ & $ 670.7 $ & $  998.5 $ & $   661.2 $ & $  556.8 $ & $   360.5 $ & $  517.0 $ & $  780.1 $ & $  952.7 $ & $ 1003.2 $  \\
$M^{3G}_{\Pl L}$[GeV]        & $  195.8 $ & $ 323.8 $ & $ 1449.6 $ & $   282.4 $ & $  394.7 $ & $   250.1 $ & $  259.7 $ & $  256.8 $ & $  352.8 $ & $  307.4 $  \\
$M^{3G}_{\Pe R}$[GeV]        & $  133.6 $ & $ 218.6 $ & $ 1438.9 $ & $   170.0 $ & $  289.5 $ & $   180.9 $ & $  230.5 $ & $  117.6 $ & $  167.2 $ & $  281.2 $  \\ \hline
\end{tabular}
\end{rotate}
\end{center}
\mycaption{\label{ta:SUSY_input} The low-energy input for the SPS scenarios. See text for details.}
\etab

\end{appendix}


\clearpage

 
\begin{thebibliography}{99}
\frenchspacing
\newcommand{\epj}[3]{{\sl Eur. Phys. J.} {\bf #1} (19#2) #3}
\newcommand{\zp}[3]{{\sl Z. Phys.} {\bf #1} (19#2) #3}
\newcommand{\np}[3]{{\sl Nucl. Phys.} {\bf #1} (19#2) #3}
\newcommand{\phm}[3]{{\sl Phil. Mag.} {\bf #1} (19#2) #3}
\newcommand{\pl}[3]{{\sl Phys. Lett.} {\bf #1} (19#2) #3}
\newcommand{\pr}[3]{{\sl Phys. Rev.} {\bf #1} (19#2) #3}
\newcommand{\prep}[3]{{\sl Phys.\ Rep.} {\bf #1} (19#2) #3}
\newcommand{\prl}[3]{{\sl Phys. Rev. Lett.} {\bf #1} (19#2) #3}
\newcommand{\prs}[3]{{\sl Proc. Roy. Soc.} {\bf #1} (19#2) #3}
\newcommand{\fp}[3]{{\sl Fortschr. Phys.} {\bf #1} (19#2) #3}
\newcommand{\cpc}[3]{{\sl Comput. Phys. Commun.} {\bf #1} (19#2) #3}
\newcommand{\ijmp}[3]{{\sl Int. J. Mod. Phys.} {\bf #1} (19#2) #3}
\newcommand{\nim}[3]{{\sl Nucl. Instr. Meth.} {\bf #1} (19#2) #3}
\newcommand{\nc}[3]{{\sl Nuovo Cimento} {\bf #1} (19#2) #3}
\newcommand{\vj}[4]{{\sl #1} {\bf #2} (19#3) #4}
\newcommand{\jcp}[3]{{\sl J. Comp. Phys.} {\bf #1} (19#2) #3}

\bibitem{Gerber:2007xk}
  C.~E.~Gerber {\it et al.}  [TeV4LHC Top and Electroweak Working Group],
  ``Tevatron-for-LHC report: Top and electroweak physics,''
  arXiv:0705.3251 [hep-ph].
    
\bibitem{Buescher:2006jm}
  V.~B\"uscher  {\it et al.} [TeV4LHC Landscape Working Group], 
  ``Tevatron-for-LHC report: Preparations for discoveries,''
  hep-ph/0608322.

\bibitem{vanNeerven:1991gh}
  W.~L.~van Neerven and E.~B.~Zijlstra,
  Nucl.\ Phys.\  B {\bf 382} (1992) 11
  [Erratum-ibid.\  B {\bf 680} (2004) 513];
  R.~V.~Harlander and W.~B.~Kilgore,
  Phys.\ Rev.\ Lett.\  {\bf 88} (2002) 201801
  [hep-ph/0201206];
  C.~Anastasiou, L.~J.~Dixon, K.~Melnikov and F.~Petriello,
  Phys.\ Rev.\ Lett.\  {\bf 91} (2003) 182002
  [hep-ph/0306192];
  C.~Anastasiou, L.~J.~Dixon, K.~Melnikov and F.~Petriello,
  Phys.\ Rev.\  D {\bf 69} (2004) 094008
  [hep-ph/0312266].

\bibitem{Moch:2005ky}
  S.~Moch and A.~Vogt,
  Phys.\ Lett.\  B {\bf 631} (2005) 48
  [hep-ph/0508265];
  E.~Laenen and L.~Magnea,
  Phys.\ Lett.\  B {\bf 632}, 270 (2006)
  [hep-ph/0508284];
  A.~Idilbi, X.~d.~Ji, J.~P.~Ma and F.~Yuan,
  Phys.\ Rev.\  D {\bf 73}, 077501 (2006)
  [hep-ph/0509294];
  V.~Ravindran and J.~Smith,
  Phys.\ Rev.\  D {\bf 76}, 114004 (2007)
  [arXiv:0708.1689 [hep-ph]].

\bibitem{Frixione:2006gn}
  S.~Frixione and B.~R.~Webber,
  hep-ph/0612272.
  
\bibitem{Balazs:1997xd}
  C.~Balazs and C.~P.~Yuan,
  Phys.\ Rev.\  D {\bf 56} (1997) 5558
  [hep-ph/9704258];
  R.~K.~Ellis and S.~Veseli,
  Nucl.\ Phys.\  B {\bf 511} (1998) 649
  [hep-ph/9706526];
  A.~Kulesza and W.~J.~Stirling,
  JHEP {\bf 0001} (2000) 016
  [hep-ph/9909271].

\bibitem{Frixione:2004us}
  S.~Frixione and M.~L.~Mangano,
  JHEP {\bf 0405} (2004) 056
  [hep-ph/0405130].

\bibitem{Zykunov:2001mn}
  V.~A.~Zykunov,
  Eur.\ Phys.\ J.\ direct C {\bf 3} (2001) 1
  [hep-ph/0107059].

\bibitem{Dittmaier:2001ay}
  S.~Dittmaier and M.~Kr\"amer,
  Phys.\ Rev.\ D {\bf 65} (2002) 073007
  [hep-ph/0109062].

\bibitem{Baur:2004ig}
  U.~Baur and D.~Wackeroth,
  Phys.\ Rev.\  D {\bf 70} (2004) 073015
  [hep-ph/0405191];
  A.~Arbuzov, D.~Bardin, S.~Bondarenko, P.~Christova, L.~Kalinovskaya,
  G.~Nanava and R.~Sadykov,
  Eur.\ Phys.\ J.\  C {\bf 46} (2006) 407
  [Erratum-ibid.\  C {\bf 50} (2007) 505]
  [hep-ph/0506110].

\bibitem{CarloniCalame:2006zq}
  C.~M.~Carloni Calame, G.~Montagna, O.~Nicrosini and A.~Vicini,
  JHEP {\bf 0612} (2006) 016
  [hep-ph/0609170].

\bibitem{Buttar:2006zd}
  C.~Buttar {\it et al.},
  ``Les Houches physics at TeV colliders 2005, standard model, QCD, EW, and
  Higgs working group: Summary report,''
  hep-ph/0604120.

\bibitem{DK_LH} S.~Dittmaier and M.~Kr\"amer, in \citere{Buttar:2006zd}.

\bibitem{Arbuzov:2007kp}
  A.~B.~Arbuzov and R.~R.~Sadykov,
  arXiv:0707.0423 [hep-ph].

\bibitem{CarloniCalame:2003ux}
  C.~M.~Carloni Calame, G.~Montagna, O.~Nicrosini and M.~Treccani,
  Phys.\ Rev.\  D {\bf 69} (2004) 037301
  [hep-ph/0303102].

\bibitem{Placzek:2003zg}
  W.~Placzek and S.~Jadach,
  Eur.\ Phys.\ J.\  C {\bf 29} (2003) 325
  [hep-ph/0302065];
  C.~M.~Carloni Calame, S.~Jadach, G.~Montagna, O.~Nicrosini and W.~Placzek,
  Acta Phys.\ Polon.\  B {\bf 35} (2004) 1643
  [hep-ph/0402235].

\bibitem{Cao:2004yy}
  Q.~H.~Cao and C.~P.~Yuan,
  Phys.\ Rev.\ Lett.\  {\bf 93} (2004) 042001
  [hep-ph/0401026];
  B.~F.~L.~Ward, C.~Glosser, S.~Jadach and S.~A.~Yost,
  Int.\ J.\ Mod.\ Phys.\  A {\bf 20} (2005) 3735
  [hep-ph/0411047];
  B.~F.~L.~Ward and S.~A.~Yost,
  Acta Phys.\ Polon.\  B {\bf 38} (2007) 2395
  [arXiv:0704.0294 [hep-ph]];
  G.~Montagna, talk given at ``Loopfest VI: Radiative Corrections for the LHC and ILC'', Fermilab, 2007.

\bibitem{Kuhn:2007qc}
  J.~H.~K\"uhn, A.~Kulesza, S.~Pozzorini and M.~Schulze,
  Phys.\ Lett.\  B {\bf 651} (2007) 160
  [hep-ph/0703283] and
%
  Nucl.\ Phys.\  B {\bf 797} (2008) 27
  [arXiv:0708.0476 [hep-ph]];
  W.~Hollik, T.~Kasprzik and B.~A.~Kniehl,
  Nucl.\ Phys.\  B {\bf 790} (2008) 138
  [arXiv:0707.2553 [hep-ph]].

\bibitem{Gounaris:2007gx}
  G.~J.~Gounaris, J.~Layssac and F.~M.~Renard,
  Phys.\ Rev.\  D {\bf 77} (2008) 013003
  [arXiv:0709.1789 [hep-ph]].

\bibitem{Dittmaier:2008md}
  S.~Dittmaier, A.~Kabelschacht and T.~Kasprzik,
  to appear in Nucl.\ Phys.\  B
  [arXiv:0802.1405 [hep-ph]].
  
\bibitem{Hahn:2000kx}
  T.~Hahn,
  Comput.\ Phys.\ Commun.\  {\bf 140} (2001) 418
  [hep-ph/0012260];
  T.~Hahn and C.~Schappacher,
  Comput.\ Phys.\ Commun.\  {\bf 143} (2002) 54
  [hep-ph/0105349].

\bibitem{Hahn:1998yk}
  T.~Hahn and M.~Perez-Victoria,
  Comput.\ Phys.\ Commun.\  {\bf 118} (1999) 153
  [hep-ph/9807565].

\bibitem{Dittmaier:1999mb}
  S.~Dittmaier,
  Nucl.\ Phys.\ B {\bf 565} (2000) 69
  [hep-ph/9904440].

\bibitem{Kinoshita:1962ur}
  T.~Kinoshita,
  J.\ Math.\ Phys.\  {\bf 3} (1962) 650;
%
  T.~D.~Lee and M.~Nauenberg,
  Phys.\ Rev.\  {\bf 133} (1964) B1549.

\bibitem{Consoli:1989fg}
  M.~Consoli, W.~Hollik and F.~Jegerlehner,
  Phys.\ Lett.\  B {\bf 227} (1989) 167.

\bibitem{Diener:2005me}
  K.~P.~Diener, S.~Dittmaier and W.~Hollik,
  Phys.\ Rev.\ D {\bf 72} (2005) 093002
  [hep-ph/0509084].

\bibitem{Fadin:1999bq}
  V.~S.~Fadin, L.~N.~Lipatov, A.~D.~Martin and M.~Melles,
  Phys.\ Rev.\  D {\bf 61} (2000) 094002
  [hep-ph/9910338].

\bibitem{Ciafaloni:2000df}
  M.~Ciafaloni, P.~Ciafaloni and D.~Comelli,
  Phys.\ Rev.\ Lett.\  {\bf 84} (2000) 4810
  [hep-ph/0001142].

\bibitem{Hori:2000tm}
  M.~Hori, H.~Kawamura and J.~Kodaira,
  Phys.\ Lett.\  B {\bf 491} (2000) 275
  [hep-ph/0007329].

\bibitem{Melles:2001dh}
  M.~Melles,
  Eur.\ Phys.\ J.\  C {\bf 24} (2002) 193
  [hep-ph/0108221].

\bibitem{Beenakker:2001kf}
  W.~Beenakker and A.~Werthenbach,
  Nucl.\ Phys.\  B {\bf 630} (2002) 3
  [hep-ph/0112030].

\bibitem{Denner:2003wi}
  A.~Denner, M.~Melles and S.~Pozzorini,
  Nucl.\ Phys.\  B {\bf 662} (2003) 299
  [hep-ph/0301241].

\bibitem{Jantzen:2005xi}
  B.~Jantzen, J.~H.~K\"uhn, A.~A.~Penin and V.~A.~Smirnov,
  Phys.\ Rev.\  D {\bf 72} (2005) 051301
  [Erratum-ibid.\  D {\bf 74} (2006) 019901]
  [hep-ph/0504111] and
%
  Nucl.\ Phys.\  B {\bf 731} (2005) 188
  [Erratum-ibid.\  B {\bf 752} (2006) 327]
  [hep-ph/0509157].

\bibitem{Denner:2006jr}
  A.~Denner, B.~Jantzen and S.~Pozzorini,
  Nucl.\ Phys.\  B {\bf 761} (2007) 1
  [hep-ph/0608326].

\bibitem{Ciafaloni:2006qu}
  P.~Ciafaloni and D.~Comelli,
  JHEP {\bf 0609} (2006) 055
  [hep-ph/0604070].

\bibitem{Baur:2006sn}
  U.~Baur,
  Phys.\ Rev.\  D {\bf 75} (2007) 013005
  [hep-ph/0611241].

\bibitem{Martin:2004dh}
  A.~D.~Martin, R.~G.~Roberts, W.~J.~Stirling and R.~S.~Thorne,
  Eur.\ Phys.\ J.\ C {\bf 39} (2005) 155
  [hep-ph/0411040].

\bibitem{Kuraev:1985hb}
  E.~A.~Kuraev and V.~S.~Fadin,
  Sov.\ J.\ Nucl.\ Phys.\ {\bf 41} (1985) 466
  [Yad.\ Fiz.\  {\bf 41} (1985) 733];
  G.~Altarelli and G.~Martinelli,
  {\it  In Ellis, J. ( Ed.), Peccei, R.d. ( Ed.): Physics At Lep, Vol. 1,
  47-57;}
  O.~Nicrosini and L.~Trentadue,
  Phys.\ Lett.\ B {\bf 196} (1987) 551;
  O.~Nicrosini and L.~Trentadue,
  Z.\ Phys.\ C {\bf 39} (1988) 479;
  F.~A.~Berends, W.~L.~van Neerven and G.~J.~H.~Burgers,
  Nucl.\ Phys.\ B {\bf 297} (1988) 429
  [Erratum-ibid.\ B {\bf 304} (1988) 921];
  A.~B.~Arbuzov,
  Phys.\ Lett.\ B {\bf 470} (1999) 252
  [hep-ph/9908361].

\bibitem{Yao:2006px}
  W.~M.~Yao {\it et al.}  [Particle Data Group],
  J.\ Phys.\ G {\bf 33} (2006) 1.

\bibitem{Jegerlehner:2001ca}
  F.~Jegerlehner,
  DESY 01-029, LC-TH-2001-035 [hep-ph/0105283].
  
\bibitem{Abazov:2003sv}
  V.~M.~Abazov {\it et al.}  [CDF Collaboration],
  Phys.\ Rev.\  D {\bf 70} (2004) 092008
  [hep-ex/0311039].
  
\bibitem{mcfm} 
  J.~Campbell and R.~K.~Ellis, MCFM -- Monte Carlo for FeMtobarn
  processes, \url{http://mcfm.fnal.gov/}.

\bibitem{Smith:1983aa}
  J.~Smith, W.~L.~van Neerven and J.~A.~M.~Vermaseren,
  Phys.\ Rev.\ Lett.\  {\bf 50} (1983) 1738.

\bibitem{Allanach:2002nj}
  B.~C.~Allanach {\it et al.},
  Eur.\ Phys.\ J.\ C {\bf 25} (2002) 113
  [eConf {\bf C010630} (2001) P125]
  [hep-ph/0202233].

\bibitem{SPShomepage} see: \url{http://www.cpt.dur.ac.uk/~georg/sps/}

\end{thebibliography}
\end{document}